\newcommand*{\AuthorVersion}{}

\ifdefined\AuthorVersion
    \documentclass[acmtog,authorversion,screen]{acmart}
    \makeatletter
    \let\@authorsaddresses\@empty
    \makeatother
  \else
    \documentclass[acmtog,screen]{acmart}
\fi

\copyrightyear{2024}
\acmYear{2024}
\setcopyright{rightsretained}
\acmConference[SA Conference Papers '24]{SIGGRAPH Asia 2024 Conference Papers}{December 3--6, 2024}{Tokyo, Japan}
\acmBooktitle{SIGGRAPH Asia 2024 Conference Papers (SA Conference Papers '24), December 3--6, 2024, Tokyo, Japan}\acmDOI{10.1145/3680528.3687636}
\acmISBN{979-8-4007-1131-2/24/12}

\input{macros}

\usepackage{hyperref}
\usepackage{cleveref}
\usepackage{subcaption}
\usepackage{calc}
\usepackage{booktabs}
\usepackage{bm}

\ifdefined\TAPSVersion
  \usepackage{acmart-taps}
\else
\fi

\usepackage[ruled]{algorithm2e} 

\SetAlFnt{\small}
\SetAlCapFnt{\small}
\SetAlCapNameFnt{\small}
\SetAlCapHSkip{0pt}
\SetKwComment{Comment}{/* }{ */}

\begin{document}
\title{MARS: Multi-sample Allocation through Russian roulette and Splitting}

\author{Joshua Meyer}
\orcid{0009-0009-8703-1136}
\affiliation{
  \institution{Saarland University}
  \department{Saarland Informatics Campus}
  \city{Saarbrücken}
  \country{Germany}
}
\email{joshua.meyer@cs.uni-saarland.de}

\author{Alexander Rath}
\orcid{0000-0002-6084-2942}
\affiliation{%
  \institution{Saarland University}
  \department{Saarland Informatics Campus}
  \city{Saarbrücken}
  \country{Germany}
}
\email{rath@cg.uni-saarland.de}

\author{Ömercan Yazici}
\orcid{0000-0003-0306-757X}
\affiliation{%
  \institution{Saarland University}
  \department{Saarland Informatics Campus}
  \city{Saarbrücken}
    \country{Germany}
}
\email{yazici@cg.uni-saarland.de}

\author{Philipp Slusallek}
\orcid{0000-0002-2189-2429}
\affiliation{
  \institution{German Research Center for Artificial Intelligence}
  \department{Saarland Informatics Campus}
  \city{Saarbrücken}
  \country{Germany}
}
\affiliation{%
  \institution{Saarland University}
  \department{Saarland Informatics Campus}
  \city{Saarbrücken}
  \country{Germany}
}
\email{philipp.slusallek@dfki.de}

\begin{abstract}
  Multiple importance sampling (MIS) is an indispensable tool in rendering that constructs robust sampling strategies by combining the respective strengths of individual distributions.Its efficiency can be greatly improved by carefully selecting the number of samples drawn from each distribution, but automating this process remains a challenging problem.Existing works are mostly limited to mixture sampling, in which only a single sample is drawn in total, and the works that do investigate multi-sample MIS only optimize the sample counts at a per-pixel level, which cannot account for variations beyond the first bounce.Recent work on Russian roulette and splitting has demonstrated how fixed-point schemes can be used to spatially vary sample counts to optimize image efficiency but is limited to choosing the same number of samples across all sampling strategies.Our work proposes a highly flexible sample allocation strategy that bridges the gap between these areas of work.We show how to iteratively optimize the sample counts to maximize the efficiency of the rendered image using a lightweight data structure, which allows us to make local and individual decisions per technique.We demonstrate the benefits of our approach in two applications, path guiding and bidirectional path tracing, in both of which we achieve consistent and substantial speedups over the respective previous state-of-the-art.
\end{abstract}

%
%

%
%


\begin{teaserfigure}
    \ifdefined\AuthorVersion
      \includegraphics[width=\textwidth,interpolate=false,clip,trim={0.265cm 0 0 0}]{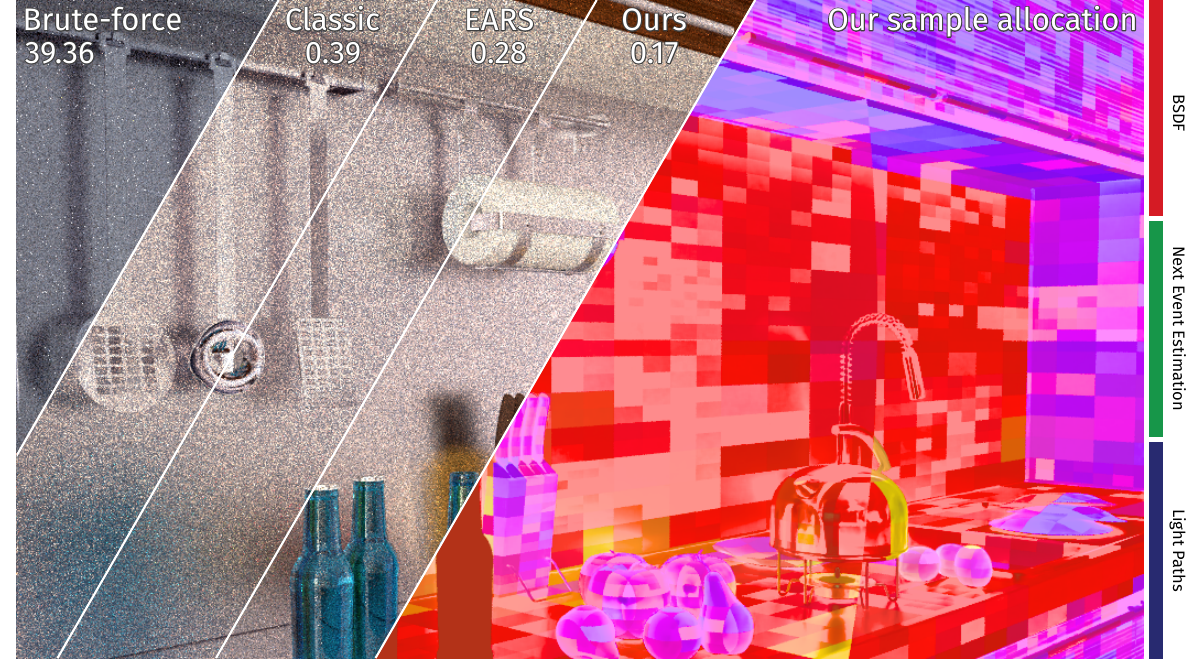}
    \else
      \includegraphics[width=\textwidth,interpolate=false,clip,trim={0.265cm 0 0 0}]{assets/TEASER-smaller.pdf}
    \fi
    \caption{
        The performance of multiple importance sampling can be greatly improved by choosing appropriate sample counts for each sampling strategy. Here, we compare the performance of our fixed-point approach applied to bidirectional path tracing (10 minute equal-time renders). Existing methods are either limited to per-pixel decisions (``brute-force'' \cite{grittmannEfficiencyawareMultipleImportance2022}) or to choosing the same amount of samples for each strategy (``EARS'' \cite{rathEARSEfficiencyawareRussian2022}). Our method learns spatially varying sample counts individually for each strategy, visualized on the right, which increases performance by $65\%$ over previous works.
    }
    \Description{Teaser figure featuring the kitchenette scene. The image is split into 5 parts. The first 4 compare the results of previous work and our approach, showcasing a significant reduction in noise levels that is visible in the image as well as in the given relative mean-squared error. The fifth part shows the local sample allocation between BSDF, NEE and light path sampling our approach decides for which is the reason for the improvement.}
    \label{fig:teaser}
\end{teaserfigure}

\maketitle

\section{Introduction}
Synthetically generating realistic images from virtual, mathematical descriptions is an essential aspect in various fields such as multimedia, gaming, and advertisement. This is achieved by reconstructing the image from a sample of admissible light paths, which are commonly obtained through unidirectional or bidirectional path tracing.

Despite decades of research, there is no single sampling distribution that robustly captures all effects. Instead, contemporary works employ a blend of multiple strategies to attain completeness, using a technique known as \emph{multiple importance sampling} (MIS). Many works have explored how to weight strategies in an MIS combination, but surprisingly little attention has been devoted to determining the optimal number of samples to assign to each technique in a (bidirectional) path tracing context. Although attempts have been made to address this question, none have provided entirely satisfactory solutions. In practice, sample counts are often steered by user intervention. 

We present a novel theory for MIS sample allocation. 
By considering the local variances and costs of samples, our approach efficiently distributes samples among various strategies, leading to significantly increased rendering efficiency. Moreover, we validate our theory by applying it to two rendering algorithms, demonstrating its practical viability.
In summary, our contributions are:
\begin{itemize}
    \item We derive a novel theory for multi-sample MIS budget allocation, based on the theory of fixed-point iterations (\cref{sec:theory})
    \item We show how to apply our theory in the popular context of path guiding, achieving noticeable speedups over previous works (\cref{sec:path_guiding})
    \item Additionally, we accelerate bidirectional path tracing with our approach and demonstrate equally substantial speedups (\cref{sec:bdpt})
\end{itemize}

Our Mitsuba \cite{jakobMitsubaRenderer2010} implementation of the two applications is publicly available at \url{https://github.com/woshicado/mars}.

\section{Previous work}\label{sec:previous_work}

\paragraph{Light transport}
Almost 40 years ago, \citet{kajiyaRenderingEquation1986} described how light is propagated within a scene
, which can succinctly be written as \rhl{the} path integral
\cite{veachRobustMonteCarlo1997}
\begin{align*}
    \Ipx = \int_{\PDomain} \fpx(\xb) \dxb,
\end{align*}
where a pixel's value $\Ipx$ is given by the integral of the contribution function $\fpx$ over the space $\PDomain$ of all light transport paths connecting the pixel px to a light source. Due to the high dimensionality, many discontinuities, and overall complexity, obtaining a general analytical solution is impractical. Therefore, Monte Carlo integration is typically used to compute this integral. \emph{Importance sampling} (IS) augments the integration process by sampling from a distribution
similar to the integrand in order to reduce variance.

\paragraph{Multiple importance sampling}
Often, there is not a single distribution that estimates the target sufficiently well over the whole domain.
\emph{Multiple importance sampling} (MIS) \cite{veachOptimallyCombiningSampling1995} can be utilized to combine several sampling distributions strategically to leverage their respective strengths.
To that end, a \emph{weighting heuristic} approximates which strategy performs well within which subdomain to weight them against each other, which has been explored in great detail by previous works \cite{veachOptimallyCombiningSampling1995,kondapaneniOptimalMultipleImportance2019}.

\paragraph{Path tracing}
Forward path tracing \cite{kajiyaRenderingEquation1986} is the primary choice for simulating light transport due to its flexibility and extensibility. Starting from the camera, a random walk through the scene is performed to construct a path sample.
At each intersection, the reflection properties (\emph{bidirectional scattering distribution function}, BSDF) are sampled to steer the random walk.
Additional samples, such as direct connections to light sources (\emph{next event estimation}, NEE), are commonly incorporated through MIS.
However, even with sophisticated NEE enhancements, like \emph{manifold next event estimation} \cite{hanikaManifoldNextEvent2015, zeltnerSpecularManifoldSampling2020}, forward path tracing alone does not suffice to efficiently render many effects.

\paragraph{Bidirectional methods}
In \emph{bidirectional path tracing} (BDPT)~\cite{veachBidirectionalEstimatorsLight1995}, an additional random walk from light sources is performed. Using MIS, it is possible to connect all pairs of intersections between the camera and light path to obtain many complete path samples
\cite{lafortuneBidirectionalPathTracing1993}.
It has been shown that connecting each camera path vertex to a fixed number of light subpath vertices can impact the rendering performance positively~\cite{popovProbabilisticConnectionsBidirectional2015, nabataResamplingawareWeightingFunctions2020}.

\begin{figure*}[t]
    \centering
    \includegraphics[width=\textwidth]{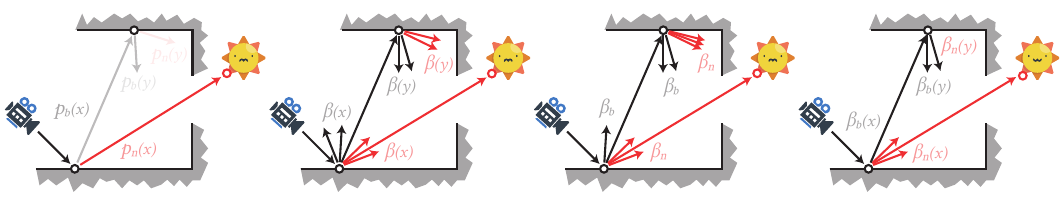}
    \captionsetup[subfigure]{aboveskip=-7pt}
    \begin{minipage}[t]{.23\textwidth}
        \subcaption{One-sample MIS}
        \label{fig:overview_os}
    \end{minipage}%
    \begin{minipage}[t]{.25\textwidth}
        \subcaption{Russian roulette and splitting}
        \label{fig:overview_rrs}
    \end{minipage}%
    \begin{minipage}[t]{.25\textwidth}
        \subcaption{Per-pixel multi-sample MIS}
        \label{fig:overview_ms}
    \end{minipage}%
    \begin{minipage}[t]{.25\textwidth}
        \subcaption{\textbf{Our approach}}
        \label{fig:overview_ours}
    \end{minipage}%
    \hfill\phantom{x}
    \caption{Existing sample allocation works share three common limitations. \textbf{(a)} One-sample MIS randomly picks a single strategy to continue the path, which can result in poor efficiency since path prefixes are not re-used. \textbf{(b)} RRS can re-use path prefixes, but chooses the same sample count for each sampling strategy independent of whether the strategy contributes or not. \textbf{(c)} Existing multi-sample MIS approaches are limited to using the same locally-unaware sample accounts along the entire path. \textbf{(d)} Our method generalizes all of these methods and chooses spatially varying sample counts for each strategy individually.}
    \Description{Visualization of the intuition of our approach versus previous work. The figure contains three subfigures featuring previous work and their shortcomings, and one subfigure showing what our approach does differently. The remaining details are contained in the figure's caption.}
    \label{fig:overview}
\end{figure*}

\paragraph{Path guiding}
Another popular approach to improve upon classical path tracing is \emph{path guiding}~\cite{jensenImportanceDrivenPath1995, vorbaOnlineLearningParametric2014, mullerPracticalPathGuiding2017}. \rhl{Using} previously acquired samples, it is possible to identify important directions and build adaptive sampling densities accordingly.
To increase robustness, guiding is jointly used with standard forward path tracing and NEE in an MIS combination.
Subsequent studies explored further improvements to increase robustness for common effects~\cite{ruppertRobustFittingParallaxaware2020,rathFocalPathGuiding2023}.

\paragraph{Mixture sampling ratios}
The ratios with which techniques are picked have a profound impact on the efficiency of MIS, but existing works on automating those share many limitations.
Most works focus on direct illumination only, optimizing the ratio between BSDF and NEE samples on a per-pixel basis, for example through
second-order approximations \cite{lu2013second} or iterative Newton-Raphson root finding \cite{sbert2019optimal,szirmay2022robust}.
\citet{murray2020learning} introduce a simple heuristic to extend these works to global illumination.
A more powerful alternative to learning ratios per pixel is to learn them per region of space. \citet{vorbaPathGuidingProduction2019} use gradient descent to compute sampling ratios between BSDF and path guiding, allowing to optimize for variance or KL divergence. \citet{rath2020variance} improved this method by considering image variance instead of local variance alone.

An important limitation shared by these approaches is that they only optimize mixture sampling (also known as \emph{one-sample MIS}), \ie the random walk in path tracing is ever only continued with a single sample.
\rhl{Consequently, they cannot take advantage of optimizing efficiency at a particular intersection without resampling the whole path prefix.}
They are also limited to optimizing efficiency locally since the ratios are optimized independently.

\paragraph{Russian roulette and splitting}
An integral component of path tracing is Russian roulette and splitting (RRS) \cite{arvoParticleTransportImage1990}.
RR entails prematurely terminating paths based on their current throughput while splitting increases the number of subpaths at a given intersection. More advanced approaches take into consideration additional metrics. \citet{vorbaAdjointdrivenRussianRoulette2016} propose a method that factors in expectations of future contributions. Further, EARS \cite{rathEARSEfficiencyawareRussian2022} employs a fixed-point iteration scheme to optimize RRS factors for efficiency, based on the theoretical optimal derivations of \citet{bolinErrorMetricMonte1997}.
While these methods can optimize for global efficiency, they only determine the total number of samples taken at each intersection and have no control over how they are allocated to individual strategies.

\paragraph{Multi-sample allocation}
In \emph{multi-sample MIS}, each distribution gets to draw an arbitrary amount of samples. This combines the strengths of mixture sampling (\ie determining the ratios across techniques) and RRS (\ie determining the total number of samples).
However, we show that combining existing works from both domains, even if they are individually optimal, is insufficient to achieve optimal performance.
Hence, they must be optimized jointly.

By modeling the expected computational cost of an MIS combination, \citet{grittmannEfficiencyawareMultipleImportance2022} globally optimize BDPT efficiency using a simple brute-force search through a set of candidate sample counts. Due to the curse of dimensionality, this technique is limited to optimizing global parameters (such as light path counts) or boolean per-pixel decisions.

Our approach (\cref{fig:overview}) similarly optimizes the efficiency of multi-sample MIS combinations but does so at a much higher granularity of optimizing individual sampling strategies with spatially varying sample counts. This is enabled through a fixed-point scheme similar to EARS \cite{rathEARSEfficiencyawareRussian2022}, which we generalize for MIS.

\section{Theory}\label{sec:theory}
In the following, we derive an efficiency-aware multi-sample MIS budget allocation strategy.
We approach this problem by formulating a general model, analyzing its properties, and deriving a fixed-point scheme.
In this section, we outline the key equations and arguments necessary to follow the idea. A full derivation can be found in the supplementary material.

\paragraph{The model.}
In multi-sample MIS, we are interested in finding an integral value $I = \int_\mathcal{X} f(x) \,\mathrm{d}x$ by combining estimates from $\nt$ different sampling strategies with varying sample counts $\sf_t$
\begin{align}
  \overallEst = \sum_{t=1}^{\nt} \frac{1}{\sf_t} \sum_{s=1}^{\sf_t} \est{I_t(x_{t,s})} =
  \sum_{t=1}^{\nt} \est{I_t(x_{t,\cdot}); \sf_t}
  , \label{eq:OverallEstimator}
\end{align}
where $\est{I; \beta}$ denotes a secondary estimator for $I$ with $\beta$ samples, and $\est{I_t}$ denotes the primary estimator of technique $t$.

An important aspect
is that estimators often capture different, possibly disjoint parts of the integrand. In path tracing, for example, we commonly have two subintegrands: direct light (estimated by NEE and BSDF sampling), and indirect light (only estimated by BSDF sampling). We address this by expressing the integrand as a sum of subintegrands $f = \sum_{i=1}^{\ni} f_i$ and perform MIS weighting thereof within the primary estimators $\est{I_t}$
\begin{align*}
  \est{I_t(x)} = \sum_{i=1}^{\ni}
  \frac{f_i(x)}{p_t(x)} w_{it}(x),
\end{align*}
where $p_t$ is the probability density function of technique $t$ and $w_{it}$~is the MIS weight with respect to all other methods that estimate integrand $f_i$.

\paragraph{Continuous sample counts.}
Instead of restricting ourselves to integer sample counts $\sf_t \in \mathbb{N}$, we shift the domain to $\mathbb{R}^{+}$ to make the problem continuous. Since, in reality, we can only perform an integer number of samples, we employ a stochastic rounding function
\begin{align*}
  r(\sf)
  =
  \begin{cases}
      \fsf + 1&\quad\text{with probability $\sf - \fsf$},\\
      \fsf&\quad\text{otherwise},
  \end{cases}
\end{align*}
where $\lfloor\cdot\rfloor$ is the floor function.
\rhl{The estimator $\est{I}$ still divides by the real-valued $\sf_t$, forming an unbiased stochastic sampling scheme.}
This is analogous to how previous works determine sample counts for RRS~\cite{vorbaAdjointdrivenRussianRoulette2016,rathEARSEfficiencyawareRussian2022}, but generalizes the concept to individual techniques instead of using the same splitting factor across all of them.

\begin{figure*}
    \centering
    \includegraphics[width=\textwidth]{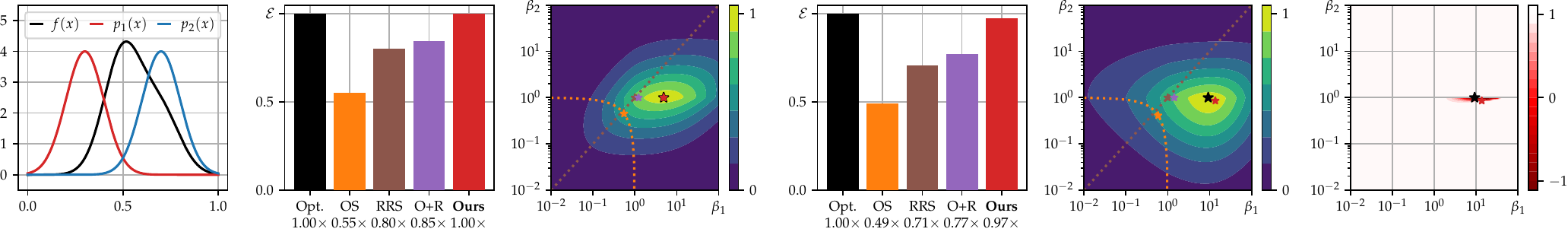}
    \captionsetup[subfigure]{aboveskip=-7pt}
    \begin{minipage}[t]{.16\textwidth}
        \subcaption{Integrand}
        \label{fig:2d_integrand}
    \end{minipage}%
    \begin{minipage}[t]{.33\textwidth}
        \subcaption{Budget-unaware balance heuristic}
        \label{fig:2d_variance}
    \end{minipage}
    \begin{minipage}[t]{.33\textwidth}
        \subcaption{Budget-aware balance heuristic}
    \label{fig:2d_rrs_factors}
    \end{minipage}
    \begin{minipage}[t]{.16\textwidth}
        \subcaption{Proxy error}
    \end{minipage}%
    \caption{
        \textbf{(a)} We evaluate the performance of our model on a simple 1D example, in which a single integrand $f$ is estimated by two techniques $p_1$ and $p_2$, the latter of which matches slightly better but is $30\times$ as expensive as $p_1$.
        \textbf{(b)} We compare the efficiency of four approaches: Optimal mixture sampling (``OS''), optimal Russian roulette and splitting (``RRS''), combining the two (``O+R''), and our proxy model (``Ours''). Note that combining optimal mixture sampling and RRS does not yield optimal performance. Our model optimizes ratios and total sample counts jointly, resulting in optimal performance for budget-unaware MIS weights.
        \textbf{(c)} For budget-aware MIS weights, the optimum predicted by our model can deviate from the true efficiency optimum.
        \textbf{(d)} We investigate the error of our proxy by plotting the dot product of the true gradients and our proxy gradients (``1'' indicates a perfect match, ``-1'' indicates opposing directions).
    }
    \Description{1D comparison to demonstrate the accuracy of our proxy, consisting of 4 subfigures. The first one shows a plot featuring the integrand and the two techniques used to estimate it. The second one contains a bar and contour plot showing that in the budget-unaware scenario our method can move freely across the whole domain and find the optimum, whereas previous work can only move on certain slices. In the third one, we show the same to demonstrate that our proxy is accurate and still close to the true optimum. The fourth subfigure shows that our proxy model closely matches the true model almost everywhere by visualizing the locations where they differ.}
    \label{fig:1DSketch}
\end{figure*}

\paragraph{The objective.}
Our goal is to assign an optimal number of samples to each strategy, i.e., to find optimal $\sf_t$.
Previous works have mostly focused on optimizing variance, which works well as long as the computational cost is largely unaffected by the choice of $\sf_t$, such as when optimizing one-sample MIS combinations.
In multi-sample MIS the cost highly depends on the number of samples, and we must therefore consider the \emph{efficiency} $\mathcal{E}$ \cite{pharr2016physically}
\begin{align*}
  \mathcal{E}\left[\est{I}\right] = \frac{1}{\nV{\est{I}}\nC{\est{I}}}.
\end{align*}
\rhl{Consequently, to optimize the efficiency, we need to predict the variance and cost for any choice of $\sf_t$.}

\paragraph{Cost model}
The cost of our multi-sample MIS estimator is given by the sum of all techniques' sample counts multiplied by the cost of taking a sample for this technique, i.e., $\nCOST[\overallEst] = \sum_{t=1}^{\nt} \sf_t \nCOST_t + \nCOST_{\Delta}$, where $\nCOST_{\Delta}$ is a constant to incorporate overhead costs.

\paragraph{Variance model}
We will assume that the techniques are uncorrelated, which allows us to express the variance of the MIS estimator as the sum of the variances of all techniques
\begin{align*}
  \nV{\overallEst}
  =
  \sum_{t=1}^{\nt} \nV{\est{I_t; \beta_t}}.
\end{align*}
We can express the variance of the secondary estimator in terms of the variance and expected value of the primary estimator:
\begin{align*}
    \nV{\est{I_t; \beta_t}}
    =
        \frac{1}{\sf_t} \nV{\est{I_t}}
        + \rho(\sf_t) \, \mathbb{E}^{2}\left[\langle I_t\rangle\right]
    ,
\end{align*}
where $\rho(\sf) = \frac{(\sf - \fsf) (\csf - \sf)}{\sf^2}$ results from stochastic rounding.

\paragraph{Optimization.}
We want to find the optimal $\sf_t$ that maximize efficiency. To reduce clutter, our derivation pursues the equivalent goal of minimizing inverse efficiency, i.e.,
$\mathcal{E}^{-1}\left[\est{I}\right]$.
A necessary condition for a minimum with respect to $\sf_t$ is that the partial derivatives of inverse efficiency equal $0$.
Since the complex shape of $\rho(\sf_t)$ makes analysis challenging, we resort to the same simplification employed by previous works \cite{rathEARSEfficiencyawareRussian2022}
\begin{equation*}
    \nV{\est{I_t; \beta_t}}
    = \begin{cases}
        \frac{1}{\sf_t} \nE{\est{I_t}\smash{^2}} -
            \nEsq{\est{I_t}}
        & \text{if}\;\sf_t \leq 1,\\
        \frac{1}{\sf_t} \nV{\est{I_t}} & \text{\rhl{otherwise}}.
    \end{cases}
\end{equation*}

This simplification only affects the $\beta > 1$ case, where it drops the noise introduced by stochastic rounding to arrive at a convex expression for variance. An analysis of this simplification, along with alternative options for rounding, can be found in the supplemental.

\paragraph{Proxy model}
Even for simple weighting functions such as the balance heuristic, the efficiency is not generally convex in $\beta_t$ due to the terms the MIS weights introduce in the variance derivatives
\begin{equation*}
    \frac{{\rm d}\mathbb{V}[\langle I\rangle]}{{\rm d}\beta_t}
=
    \begin{cases}
        - \frac{1}{\sf_t^2}
\mathbb{E}[\langle I_t\rangle^{2}]
+ \sum_{k=1}^{\nt} \frac{1}{\sf_k}
\frac{{\rm d}\mathbb{E}[\langle I_k\rangle^2]}{{\rm d}\beta_t}
        & \text{if}\;\sf_t \leq 1,\\
        - \frac{1}{\sf_t^2} \mathbb{V}[\langle I_t\rangle] + \sum_{k=1}^{\nt} \frac{1}{\sf_k}
\frac{{\rm d}\mathbb{V}[\langle I_k\rangle]}{{\rm d}\beta_t}
        & \text{otherwise}.
    \end{cases}
\end{equation*}
We instead resort to optimizing a proxy for efficiency, in which the MIS weight dependencies are dropped from the variance derivative
\begin{align*}
\frac{{\rm d}\mathbb{V}[\langle I\rangle]}{{\rm d}\beta_t}
&=
    \begin{cases}
        - \frac{1}{\sf_t^2}
\mathbb{E}[\langle I_t\rangle^{2}]
        & \text{if}\;\sf_t \leq 1,\\
        - \frac{1}{\sf_t^2}
\mathbb{V}[\langle I_t\rangle]
        & \text{otherwise}.
    \end{cases}
\end{align*}
This proxy is optimal for \emph{budget-unaware} MIS weights (\ie weights independent of $\sf_t$)%
, but for \emph{budget-aware} weights
\rhl{(\ie weights containing the techniques' budgets $\sf_t$)}
its optimum can deviate from the true optimal solution.
We perform two evaluations to verify the effectiveness of this proxy: First, we compare its derivatives and optima to the true efficiency for simple one-dimensional functions~(see \cref{fig:1DSketch} and further examples in the supplemental). Second, in our two rendering applications, we demonstrate that budget-aware weights with this proxy robustly outperform theoretically optimal allocations of budget-unaware weights.

\paragraph{Fixed point}
The optimal $\sf_t$ that minimize inverse efficiency in our proxy model are given by
\begin{align}\label{eq:optimalBeta}
  \sf_t =
    \begin{cases}
        \sf_t^\text{RR} = \sqrt{
            \frac{\nCOST[\overallEst]}{\nCOST_t}
            \frac{\nE{\est{I_t}^2}}{\nV{\overallEst}}
        } &\text{if $\sf_t^\text{RR} < 1$},\\
        \sf_t^\text{\parbox{\widthof{RR}}S} = \sqrt{
            \frac{\nCOST[\overallEst]}{\nCOST_t}
            \frac{\nV{\est{I_t}}}{\nV{\overallEst}}
        } &\text{if $\sf_t^\text{\parbox{\widthof{RR}}S} > 1$},\\
        1 &\text{\rhl{otherwise}} .
    \end{cases}
\end{align}

\begin{figure}
    \centering
    \includegraphics[width=\columnwidth]{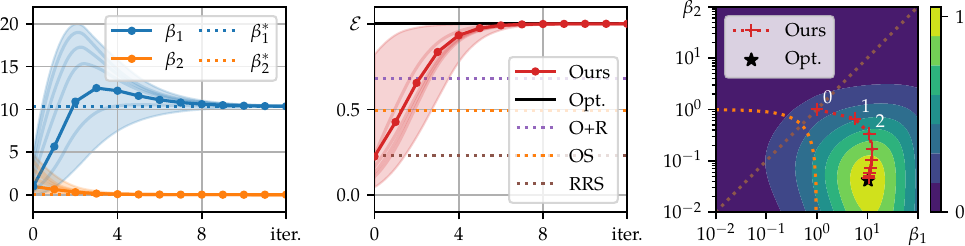}
    \captionsetup[subfigure]{aboveskip=-7pt}
    \begin{minipage}[t]{.34\columnwidth}
        \subcaption{Sample budgets}
    \end{minipage}%
    \begin{minipage}[t]{.34\columnwidth}
        \subcaption{Efficiency}
    \end{minipage}
    \begin{minipage}[t]{.31\columnwidth}
        \subcaption{Landscape}
    \end{minipage}
    \caption{\textbf{(a)}~We visualize the quick convergence of the sample allocations $\sf_1$ and $\sf_2$ towards their optimal values for a simple 1D example. The shaded area corresponds to trajectories of various initializations.
    \textbf{(b)}~After 5 iterations we are close to the optimal efficiency. While training longer brings us closer, it yields diminishing returns.
    \textbf{(c)}~We show the path our fixed-point iteration takes along the efficiency landscape.
    }
    \Description{2 Line graphs showing the number of samples and overall efficiency converging toward the optimal values. The values are negligibly close to the optimum in as few as 6 iterations. A third plot shows the efficiency landscape as a contour plot and the trajectory over the fixed-point iteration, leading to the optimal allocation for efficiency.}
    \label{fig:1DConvergence}
\end{figure}

An analytical solution to this equation is impractical because $\sf_t$ appears non-linearly on the right-hand side as part of $\nCOST[\overallEst]$ and $\nV{\overallEst}$.
Therefore, to find the root numerically, we employ a fixed-point iteration scheme similar to the work of \citet{rathEARSEfficiencyawareRussian2022}. Our equation is of the same form and, thus, we can iteratively approach the root by using \cref{eq:optimalBeta} as an update function.
In practice, this converges quickly after only a few iterations as depicted in \cref{fig:1DConvergence}.

\subsection{Application to Path Tracing}
We will now discuss how to apply our theory to path tracing, for which we extend our theory to multiple spatially varying estimators.

\paragraph{Primary estimator}
In path tracing, we combine multiple strategies for estimating reflected radiance $L_r$ at each intersection
\begin{equation*}
    \est{L_{\text{r},t}(x, \omega_{\rm o})} =
    \frac{
        \est{L_\text{i}(x, \omega_{\rm i})} \,
        B(x, \omega_{\rm i}, \omega_{\rm o}) \,
        \abs{\cos \theta_{\rm i}}
    }{
        p_t(\omega_{\rm i} \mid \rhl{x}, \omega_{\rm o})
    } \,
    w_{t}(\rhl{x}, \omega_{\rm i}, \omega_{\rm o}) ,
\end{equation*}
where technique $t$ samples $\omega_{\rm i} \sim p_t$ and $B$ denotes the BSDF.

\paragraph{Secondary estimator}
Analogous to \cref{eq:OverallEstimator}, we define the multi-sample estimator for $L_\text{r}$
\begin{equation*}
    \est{L_\text{r}(x, \omega_{\rm o})}
    =
    \rhl{
    \sum_{t=1}^{\nt} \frac{1}{\sf_t(\xb)} \sum_{s=1}^{\sf_t(\xb)} \est{L_{\text{r},t}(x_s, \omega_{\rm o})}
    }
    =
    \sum_{t=1}^{\nt} \est{\rhl{L_{\text{r},t}}; \sf_t(\xb)} ,
\end{equation*}
where the sample budgets $\sf_t$ are optimized for each individual technique $t$ and path prefix $\xb$.
Instead of budgets that are constant across the image, we \rhl{thereby adapt them locally}.

\paragraph{Random walks}
Our random walk for a pixel $\px$ begins with a ray sampled by the camera and then recursively applies the multi-sample estimator of $L_\text{r}$ at each intersection
\begin{equation*}
    \est{\Ipx} =
    \frac{
        W_\px(x_0, \omega_{\rm i,0})
    }{
        p(x_0, \omega_{\rm i,0})
    }
    \est{L_\text{r}(x_1, \omega_{\rm o,1})}
    .
\end{equation*}

\paragraph{Multiple estimators}
Optimizing each pixel individually is insufficient to achieve a global optimum \rhl{as this will oversample bright regions}. Hence, we maximize the efficiency built from the average variance $\mathbb{V}_\text{I}$ and average cost $\mathbb{C}_\text{I}$ over the entire image instead \cite{rathEARSEfficiencyawareRussian2022}.
To not assign too much weight to bright pixels, we similarly use the \emph{relative} variance of each pixel, i.e., we divide by the pixel's ground truth $\Ipx$
\begin{align*}
  \mathcal{E}^{-1}
  =
   \underbrace{
    \left( \frac{1}{\NumPixels} \sum_{\px}^{\NumPixels} \frac{\nV{\est{\Ipx}}}{\Ipx^2} \right)
    }_{\mathbb{V}_\text{I}}
    \underbrace{
   \left( \frac{1}{\NumPixels} \sum_{\px}^{\NumPixels} \nCOST[\est{\Ipx}] \right)
   }_{\mathbb{C}_\text{I}}.
\end{align*}
\rhl{Since, in practice, $\Ipx$ is not readily available, we discuss an approximation thereof in in \cref{sec:guiding_implementation}}.

\paragraph{Optimization}
Extended to multiple spatially varying estimators, our update function becomes
\begin{align*}
    \sf_{t}(\xb) =
    \begin{cases}
        \sf_t^\text{RR} =
        \frac{T(\xb)}{\Ipx}
        \sqrt{
            \frac{\mathbb{C}_\text{I}}{\nCOST_t}
            \frac{\nE{\est{L_{\text{r},t}}^2}}{\mathbb{V}_\text{I}}
        } &\text{if $\sf_t^\text{RR} < 1$},\\
        \sf_t^\text{\parbox{\widthof{RR}}S} =
        \frac{T(\xb)}{\Ipx}
        \sqrt{
            \frac{\mathbb{C}_\text{I}}{\nCOST_t}
            \frac{\nV{\est{L_{\text{r},t}}}}{\mathbb{V}_\text{I}}
        } &\text{if $\sf_t^\text{\parbox{\widthof{RR}}S} > 1$},\\
        1 &\text{\rhl{otherwise}}.
    \end{cases}
\end{align*}
Here, $T(\xb_k)$ denotes the throughput weight of the path prefix
\begin{align*}
    T(\xb_k) =
    \frac{
        W_{\px}(x_0,\omega_{{\rm i},0})
    }{
        p(x_0,\omega_{{\rm i},0})
    }
    \prod_{j=1}^{k-1}
        \frac{
            B(x_j, \omega_{{\rm i},j}, \omega_{{\rm o},j})
            \,
            \abs{\cos \theta_{{\rm i},j}}
        }{
            p_{t_j}(\omega_{{\rm i},j} \mid x_j, \omega_{{\rm o},j})
        }
        \frac{
            w_{\text{ind},{t_j}}(\xb_j)
        }{
            \beta_{t_j}(\xb_j)
        },
\end{align*}
where $t_j$ denotes the technique that was used at the $j$-th bounce, and $w_\text{ind}$ is the MIS weight over all techniques that participate in estimating indirect light (\ie lead to a secondary estimator being performed at $x_j$).

\section{Application I: Path guiding}\label{sec:path_guiding}
We begin by demonstrating how our technique can be used in unidirectional rendering to combine three common strategies: BSDF sampling, next event estimation (NEE) and guided sampling.

\subsection{Implementation}\label{sec:guiding_implementation}
We test our method in a guided path tracer on top of the implementation of \citet{mullerPracticalPathGuiding2017} in Mitsuba \cite{jakobMitsubaRenderer2010}.
Each completed training iteration of the guiding iteration marks one step of our fixed-point scheme.
To compute the necessary variance and cost estimates, we build Monte Carlo estimates from the samples obtained throughout the iteration.

\paragraph{Local estimates}
We share the spatio-directional tree used by guiding to store the local estimates needed by our method, such as cost, variances, and second moments for each region and each technique.
For every cache $c$ and technique $t$, the first and second moments are estimated from the iteration's training samples that landed in the cache and were produced by the technique itself
\begin{align*}
    \nE{\est{L_{\text{r},t}}} \approx
    \frac{1}{n_{t,c}} \sum_{s=1}^{n_{t,c}} \est{L_{\text{r},t_s}(x_s)}
    ;\;
    \nE{\est{L_{\text{r},t}}^2} \approx
    \frac{1}{n_{t,c}} \sum_{s=1}^{n_{t,c}} \est{L_{\text{r},t_s}(x_s)}^2,
\end{align*}
from which we can compute the estimate for the variance.

To estimate the cost of each technique, we employ the well-established heuristic of using the average path length generated by the technique as a proxy for cost \cite{veachRobustMonteCarlo1997, rathEARSEfficiencyawareRussian2022}.

\paragraph{Image estimates}
In lieu of the real ground truth, we apply a denoiser \cite{oidn} to determine the pixel estimates $\Ipx$. The pixel estimates are also used to estimate the image variance
\begin{align*}
    \nV{\overallEst}
    \approx
    \frac{1}{\NumPixels} \sum_{\px}
    \frac{1}{\NumSpp} \sum_{s=1}^{\NumSpp}
    \left( \frac{\est{\Ipx}_{s} - \denIpx}{\denIpx} \right)^2 .
\end{align*}
The cost of the entire image is similarly estimated by counting the total number of rays that were traced.

\paragraph{Fixed-point iteration}
Implementing our fixed-point iteration algorithm requires an iterative rendering process, in which each iteration performs some number of samples to estimate the variances and costs, thus, allowing us to perform steps toward the fixed-point. Path guiding already uses an iterative scheme to learn its guided sampling distribution in a training phase that we can integrate the fixed-point iteration into.
We use ordinary throughput-based Russian roulette without splitting for the first three training iterations to give the data structure time to adapt and let the estimates become less noisy before we switch to our sample allocation.

\paragraph{Outlier clamping}
Outliers can drastically reduce the quality of the variance estimates. To combat this, we clamp the contribution of paths to a fixed multiple of the pixel estimate ($50$). Note that this is only used for our local- and image variance estimates, and is not used for the render itself. Similar to previous works, we also clamp the maximal number of samples per technique (to $[0.05, 20]$).

\paragraph{Handling colors}
To extend our theory to color, we perform computations channel-wise and use the $L_2$ norm to compute the final sample counts analogous to previous works \cite{rathEARSEfficiencyawareRussian2022}.


\subsection{Evaluation}

We compare our approach against the state-of-the-art in path guiding, based on mixture optimization using gradient descent \cite{vorbaPathGuidingProduction2019} in conjunction with EARS \cite{rathEARSEfficiencyawareRussian2022} to steer sample counts (``EARS'').
We use the variant that minimizes KL divergence of image contributions \cite{rath2020variance}, which works best across all our scenes.
As a baseline, \rhl{we include path guiding} with throughput-based RR (``classic RR'') starting at the 5th bounce, while all other methods start performing RR with the first bounce.

We provide results of our strategy for the budget-unaware and the budget-aware MIS weight scenarios.
\rhl{As discussed, the budget-aware configuration uses the the budget-unaware derivations as an approximation.}
An ablation (``Ours + Grad. Descent'') combines gradient descent (mixing BSDF and guiding) and our approach (sample counts of NEE and the mixture) to demonstrate that our approach chooses better mixtures than the previous state-of-the-art optimizer.

\paragraph{Setup}
All images are rendered for 5 minutes (2.5min training, 2.5min rendering) on an AMD Ryzen Threadripper 2950X.
The maximum path length is 40 and
we perform 9 training iterations whose time increases exponentially \cite{mullerPracticalPathGuiding2017}.
\rhl{We limit the data structure's footprint to a maximum of 72MB.}
The noise is quantified using relative mean squared error (relMSE), with $0.01\%$ of the brightest outliers discarded to arrive at a robust error estimate.

\begin{figure*}[p]
    \includegraphics[width=.951\textwidth,interpolate=false]{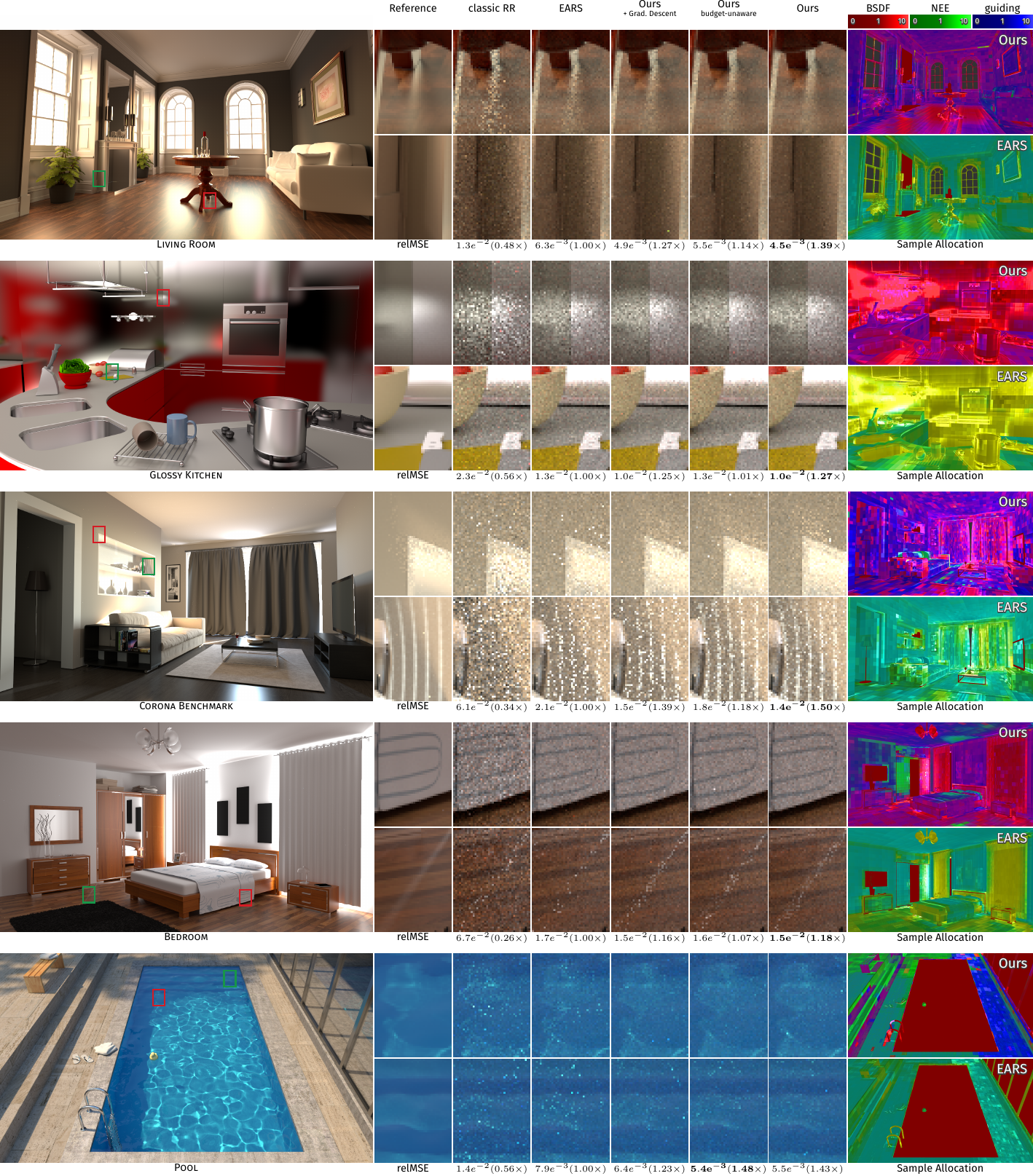}
    \caption{
        We compare our method against state-of-the-art EARS applied to guiding in 5 scenes in a 5 minute equal-time comparison. The numbers below the crops correspond to the relative mean squared error \rhl{over the whole image} (relMSE, lower is better), with the speedup compared against EARS in parentheses (higher is better). To the right, we visualize the allocation decisions EARS and our method take. The color channels red, green and blue correspond to BSDF, NEE, and guiding samples, respectively, with brighter colors indicating higher sample counts. The other 5 scenes we tested are provided in the supplementary material.
        Our ``+~Grad.~Descent'' test
        uses gradient descent for the BSDF/guiding mixture and our method to combine it with NEE using budget-aware weights.
    }
    \label{fig:GEARS-evaluation}
    \Description{Fully described in the text.}
\end{figure*}

\paragraph{Results}
We tested our method on a set of 10 common benchmark scenes, where it achieves an average speedup of 1.32$\times$ over the previous state-of-the-art\rhl{, even though it traces 17\% fewer rays}. We find that the budget-aware variant almost always yields superior performance. A selection of interesting scenes can be found in \cref{fig:GEARS-evaluation}.

In \name{Living Room}, our method automatically detects that BSDF sampling performs best to estimate the incident light on the floor. On diffuse surfaces, guiding is used instead. NEE is left mostly disabled, saving computational cost and yielding a $39\%$ speedup.

The \name{Glossy Kitchen} contains various glossy surfaces, which make guiding and NEE poor choices.
By disabling NEE almost entirely and using guiding only on the counter in the lower part where it is beneficial, we save precious time that can be spent elsewhere.

\name{Corona Benchmark} features strong indirect illumination from a directly illuminated spot on the sofa. Our method correctly identifies that any path arriving at the sofa should attempt a high number of NEE samples. Note that this would not be possible with per-pixel-based approaches, which are limited to using the same sampling ratios independent of where the path lands at later bounces.

The \name{Bedroom} is a scene where all three available techniques are useful in different areas. The left wall benefits greatly from performing NEE and every path encountering it can utilize this.

As an outdoor scene, we include the \name{Pool}. In this scene, containing many caustics, our budget-aware method achieves a speedup of~43\%. Our method correctly identifies that guiding is necessary to estimate the caustic cast by the windows on the right of the pool.

\paragraph{Convergence in practice}
\rhl{We verified} that our method converges in practice by ensuring repeated rendering yields the same results, and that the curves of average budget allocation per technique as well as overall efficiency flatten over time.
We provide these convergence plots
in the supplementary material.

\section{Application II: Bidirectional Path Tracing}\label{sec:bdpt}

In our bidirectional application, the techniques to combine are path tracing, connections with light paths, and NEE.
While NEE is a special case of light paths, its different variance and cost properties make it beneficial to assign it its own budget.
In theory, performance could be improved further by assigning separate budgets of light paths for any length, but we leave exploring this as future work.

While it might make sense to perform splitting for both the camera- and light paths, doing so comes with its own set of challenges.
Most importantly, splitting paths on both ends and combining all acquired intersections leads to an exponential increase in connections between the paths which is computationally expensive.
For this reason, we resort to throughput-based RR for the light path.

We leave exploring more sophisticated variants of bidirectional rendering, such as vertex merging, light path caches, or stochastic connections as future work.


\subsection{Implementation}
The implementation of our second application is analogous to the first.
Local estimates are recorded in the octree data structure proposed by EARS \cite{rathEARSEfficiencyawareRussian2022}.
To execute our fixed-point scheme, we rewrite Mitsuba's BDPT implementation to become progressive: Rendering occurs in short iterations that grow in duration over time, with the final image being the inverse-variance weighted combination of all iterations \cite{vorbaPathGuidingProduction2019}.
We apply a similar outlier rejection to EARS \cite{rathEARSEfficiencyawareRussian2022}, which eliminates the 10 brightest pixels in each pass for the image variance estimate.
Similar to guiding, we additionally tried applying an outlier rejection to the local samples recorded in the octree but found that this results on average in a 6\% performance decrease on the evaluated scenes: Since in this application sampling is not learned adaptively, it seems beneficial to perform excessive splitting in the presence of outliers.


\subsection{Evaluation}
We compare against EARS applied to camera paths (``EARS'') and provide the results of a vanilla BDPT baseline (``classic RR'').
Sample allocation/RRS is only performed on the camera path side, and throughput-based RR is used to construct light paths.
Additionally, we implemented a variation of efficiency-aware MIS \cite{grittmannEfficiencyawareMultipleImportance2022} (``brute-force''), which
is applied to pure bidirectional path tracing without merging and makes decisions on a per-pixel basis instead of a single global one to allocate the budget.

All images are rendered for 5 minutes on an AMD Ryzen Threadripper 2950X with 16 cores with a maximum path length of 40.
The first 3 iterations use classic RR also for the camera path to let estimates converge.
\rhl{The data structure's memory size is limited to at most 72MB.}
As in guiding, we evaluate a version with budget-unaware MIS weights which performs 13\% worse on average.
Averaged, we achieve a speedup of 1.6$\times$ over EARS applied to BDPT, \rhl{while tracing 13\% fewer rays,}
and a speedup of 2.6$\times$ over BDPT.

The teaser (\cref{fig:teaser}) shows a modified version of the \name{Kitchenette} including a small tealight.
Plain RRS cannot separate the contributions of the different techniques, wasting compute time on negligible contributions, while our method can locally identify regions where each particular sampling technique performs well.

\begin{figure*}[p]
    \includegraphics[width=.951\textwidth,interpolate=false]{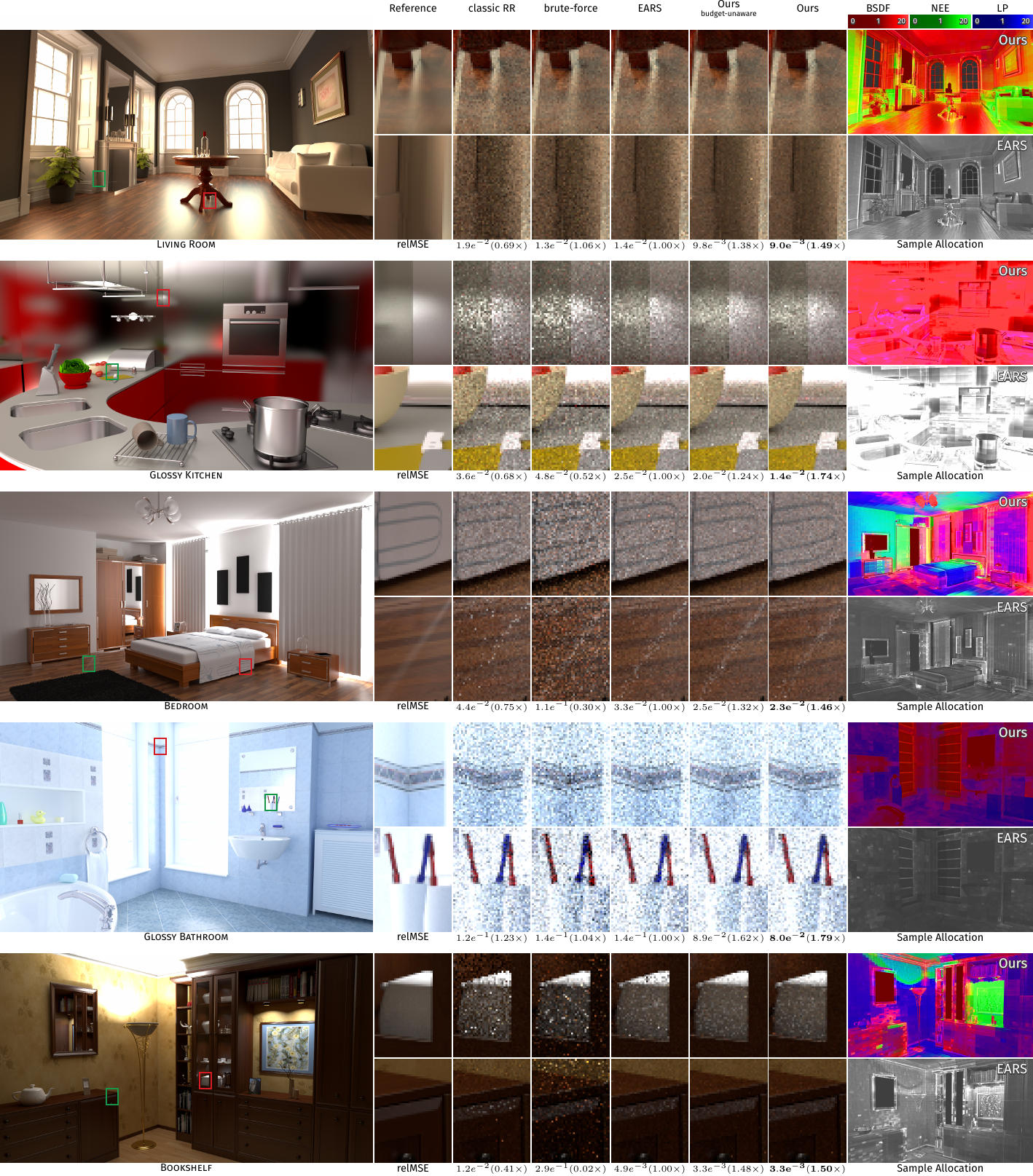}
    \caption{
        We compare our method against state-of-the-art EARS applied to BDPT in 5 scenes in a 5 minute equal-time comparison. The numbers below the crops correspond to the relative mean squared error \rhl{over the whole image} (relMSE, lower is better), with the speedup compared against EARS in parentheses (higher is better). To the right, we visualize the allocation decisions EARS and our method take. The color channels red, green and blue correspond to BSDF, NEE, and light path samples, respectively, with brighter colors indicating higher sample counts. The other 5 scenes we tested are provided in the supplementary material.
    }
    \label{fig:BEARS-evaluation}
    \Description{Fully described in the text.}
\end{figure*}

\Cref{fig:BEARS-evaluation} showcases a variety of interesting scenes, e.g., the \name{Glossy Kitchen} that shows a similar pattern. The glossy features neither benefit greatly from light connections nor NEE. Being able to automatically detect and tune down these techniques locally, still using connections on diffuse surfaces, yields a speedup of 74\%.

The \name{Glossy Bathroom} similarly avoids NEE, but for a different reason: No light is directly visible in this scene. Instead, our method relies on BSDF sampling for glossy surfaces and light paths for rougher surfaces. Similar to EARS, our method uses less splitting overall (indicated by the darker colors). EARS exhibits the same issue as in its original paper due to issues with variance estimates on glossy surfaces. While our method fundamentally suffers from the same issue, this is offset by an overall better sample allocation resulting in a 79\% speedup over EARS.

In contrast to the previous scenes, the \name{Living Room} is a scene where NEE is invaluable, given that guiding is unavailable. The sofa on the right is directly hit by the light sources and well estimated by NEE. Random BSDF bounces from diffuse elements in the scene ending at this location can exploit this effective technique. Light paths are avoided entirely, as they struggle to penetrate the room.

The \name{Bedroom} and the \name{Bookshelf} both feature difficult indirect illumination where light is only induced from small areas: from behind the curtains and from small light sources, respectively. For BSDF samples alone, it is hard to find a contributing path without the assistance of NEE or light path connections. Therefore, at intersections where there are no occluders and a direct light connection is possible, our method automatically increases the NEE sampling budget to reduce the variance at this intersection. At other intersections, with an occluder present, light connections can provide help avoiding it. Making these decisions locally in space allows for superior estimations of indirect illumination while previous work is limited to taking a single decision for the whole path.

\begin{figure}
    \centering
    \includegraphics[width=\columnwidth,clip,trim={0 0.05cm 0 0}]{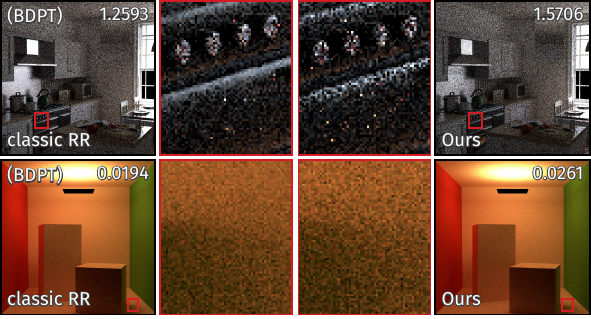}
    \caption{
        We show two failure cases. The \name{Country Kitchen} is a noisy scene in which our method requires time to converge and convert the overhead into an advantage. The 30 seconds it was rendered for \rhl{do} not suffice, but by rendering only slightly longer we surpass BDPT's performance. In the second case, we present a \name{Cornell Box}. In trivial scenes, the overhead we introduce dominates the advantage. We give the relMSE in the upper right.
    }
    \label{fig:failure}%
    \Description{Fully described in the text.}%
\end{figure}

\paragraph{Convergence in practice}
As in path guiding, \rhl{we verified} that repeated runs result in the same performance. We similarly provide plots for the convergence of average sample allocations across techniques and overall efficiency in the supplementary material.

\section{Limitations and Future Work}\label{sec:limitations}
Our method efficiently allocates sampling budgets while remaining flexible and extensible, but relies heavily on the accuracy and robustness of local estimates for variance.

\paragraph{Overhead}
Our method introduces the overhead of collecting local estimates which does not always pay off. In the presence of very short rendering times, the collected estimates have not converged enough yet when the rendering finishes. In \cref{fig:failure} we present the \name{Country Kitchen} where a rendering time of 30 seconds does not suffice, even though we outperform BDPT quickly afterward.

\paragraph{Local estimates}
Outliers in the local estimates can reduce performance due to excessive splitting. We used a simple outlier rejection scheme \rhl{in the guiding application, 
but future work could look into applying more sophisticated techniques}~\cite{zirr2018re}.

\paragraph{Discontinuity artefacts}
Noise in the spatial caches can lead to noticeable discontinuities \cite{rathEARSEfficiencyawareRussian2022}, which is especially pronounced with shorter render times. Possible mitigations could be to apply spatial denoising to the local estimates or interpolate them to arrive at smoothly varying sample counts.

\paragraph{Proxy accuracy}
Our proxy is only optimal for budget-unaware MIS weights. While it performs robustly for budget-aware weights across all our tested scenes, a more sophisticated proxy can likely increase efficiency even further.
Our supplemental includes additional 1D examples that investigate the error of the proxy further.

\paragraph{Dynamic scenes}
\rhl{We only} considered static scenes. Future work could \rhl{extend} our method to dynamic scenes, such as real-time rendering%
, in which local estimates or sample allocations from earlier frames could be re-used to kickstart the fixed-point scheme.

\paragraph{MIS weights}
We limited our analysis to the most common MIS weighting function, the balance heuristic \cite{veachOptimallyCombiningSampling1995}. Future work could investigate the impact of more sophisticated budget-aware weights on our proxy model, such as the optimal weights proposed by \citet{kondapaneniOptimalMultipleImportance2019}.

\rhl{
\paragraph{Correlated Techniques}
Our decomposition of the variance assumes that techniques are uncorrelated.
To perform optimally for correlated techniques (such as photon mapping), one also needs to track correlation statistics, which we leave as future work.
}

\paragraph{Applications outside of rendering}
\rhl{We presented our approach in the context of rendering, but it could be applied to any Monte Carlo integration problem} that relies on multi-sample MIS estimation.

\section{Conclusion}\label{sec:conclusion}
We present a novel theory on multi-sample MIS allocation strategy based on fixed-point schemes.
For budget-unaware weights, our method is proven to yield optimal performance as long as variances and costs are known accurately.
We demonstrate that our approach achieves consistent speedups over previous approaches, even with noisy estimates and budget-aware weights, in both path guiding and bidirectional path tracing.
Our theory can be applied to any Monte Carlo integration problem, even outside of rendering, in which multiple sampling techniques are combined through MIS.

\defcitealias{resources16}{Benedikt Bitterli [2016]}
\begin{acks}
This project has received funding from Deutsche Forschungsgemeinschaft (DFG, German Research Foundation) through GRK 2853/1 “Neuroexplicit Models of Language, Vision, and Action” (project number 471607914), as well as from \grantsponsor{VELUX}{Velux Stiftung}{https://veluxstiftung.ch} project~\grantnum{VELUX}{1350}.
We thank the anonymous reviewers for their insightful remarks, as well as the artists of our test scenes:
\citetalias{resources16}, Miika Aittala, Samuli Laine, and Jaakko Lehtinen (\textsc{Veach Door}), Evermotion and Tiziano Portenier (\textsc{Bookshelf}, \textsc{Glossy Bathroom}, \textsc{Glossy Kitchen}), Ondřej Karlík (\textsc{Pool}), Wig42 (\textsc{Modern Living Room}), Jay-Artist (\textsc{Country Kitchen}), and Ludvík Koutný (\textsc{Living Room}).
\end{acks}

\bibliographystyle{ACM-Reference-Format}
\balance
\bibliography{references}%
\clearpage

\ifdefined\AuthorVersion
  \clearpage
  \appendix
  \section{Derivation}
In this section, we provide the formal derivation of the fixed-point iteration scheme that is described in the paper.
We start by introducing the model and follow up with the derivation of the formulation for variance. With these foundations, we can give an expression for efficiency to take the derivative of and compute its root.

\subsection{Base Model}
We are interested in finding the value of $I=\int_{\XDomain}f(x)\dx$ by combining estimates from $\nt$ different strategies with varying sample counts $\sf_t$, leading us to a multi-sample MIS formulation
\begin{align*}
  \overallEst = \sum_{t=1}^{\nt} \frac{1}{\sf_t} \sum_{s=1}^{\sf_t} \est{I_t(x_{t,s})} =
  \sum_{t=1}^{\nt} \est{I_t(x_{t,\cdot}); \sf_t},
\end{align*}
where $\est{I; \sf}$ is a secondary estimator with sample count $\sf$. Further, we express the integrand $f$ as a sum of subintegrands $f = \sum_{i=1}^{\ni} f_i$, allowing us to consider and combine techniques that estimate different, possibly disjoint, parts of the main integrand.
This gives each primary estimator the form
\begin{align*}
  \est{I_t(x)} = \sum_{i=1}^{\ni}
  \frac{f_i(x)}{p_t(x)} w_{it}(x),
\end{align*}
where $p_t$ is the probability density function of technique $t$ and $w_{it}(x)$ is the MIS weight with respect to all other methods that estimate integrand $f_i$. Its dependency on $i$ is due to the characteristic function $\1_{it}$ in the MIS weights, which is $1$ whenever technique $t$ estimates integrand $f_i$ and $0$ otherwise. This ensures correctness and $\sum_{t=1}^{\nt} w_{it}(x) = 1$ for all $i$. For example, for the budget-unaware balance heuristic we obtain
\begin{align*}
  w_{it}(x) = \frac{\1_{it}p_{t}(x)}{\sum_{k=1}^{\nt}\1_{ik}p_{k}(x)}.
\end{align*}
For brevity, we will drop the argument of $I_t$ where it is not the subject of interest.
Note, that we can also drop the argument with explicit subscript $s$ for the $s$-th sample since the samples are pairwise independent.

\paragraph{Continuity}
Sample allocation is a discrete problem since we can only take a natural number of samples. Similar to previous work \cite{vorbaAdjointdrivenRussianRoulette2016}, we employ the stochastic rounding function
\begin{align*}
  r(\sf)
  =
  \begin{cases}
      \fsf + 1&\quad\text{with probability $\sf - \fsf$}\\
      \fsf&\quad\text{otherwise,}
  \end{cases}
\end{align*}
to allow for real-valued $\sf$ and make the optimization simpler.

\paragraph{Unbiasedness}
Having adapted the standard multi-sample MIS model and introduced stochastic rounding, we have to verify that our model still yields an unbiased estimate of $I$. This is a straightforward computation in which we use that the sum of the MIS weights over all techniques is 1:
\begingroup
\allowdisplaybreaks
\begin{align*}
    &\quad\E{\est{I}}
=
    \frac{1}{\sf_t}
    \sum_{t=1}^{\nt}
    \E{
        \sum_{s=1}^{r(\sf_t)} \est{I_t}
    }
\\=&
    \left(
        \frac{(\sf_t - \lfloor\sf_t\rfloor)(\lfloor\sf_t\rfloor+1)}{\sf_t}
        + \frac{(1 - \sf_t + \lfloor\sf_t\rfloor)\lfloor\sf_t\rfloor}{\sf_t}
    \right)
    \sum_{t=1}^{\nt}
    \E{
        \est{I_t}
    }
\end{align*}
\begin{align*}
  =&
    \sum_{t=1}^{\nt}
    \E{ \est{I_t} }
    =
    \sum_{t=1}^{\nt}
    \sum_{i=1}^{\ni}
    \E{ \frac{f_i(x)}{p_t(x)} w_{it}(x) }
\\=&
    \int_{\XDomain}
    \sum_{i=1}^{\ni}
    f_i(x)
    \underbrace{{\left(\sum_{t=1}^{\nt} w_{it}(x) \right)}}_{=1}
    \dx
    = I.
\end{align*}
\endgroup

\paragraph{Variance.}
The efficiency $\mathcal{E}$ \cite{pharr2016physically} of an estimator $\est{I}$ is given by
\begin{align*}
  \mathcal{E}[\est{I}] = \frac{1}{\V{\est{I}}\C{\est{I}}}.
\end{align*}
Consequently, to allocate sampling budgets maximizing efficiency, having formulations for variance and cost is essential. We start with variance. It is generally given by
\begin{align*}
    \V{\est{I}} = \E{\est{I}^2} - \Esq{\est{I}}.
\end{align*}

We already know that $\Esq{\est{I}} = I^2$ since our estimator is unbiased. Thus, we only need to compute the second moment of our estimator
\begin{align*}
    \E{\est{I}^2}
&=
    \E{\left(\sum_{t=1}^{\nt} \est{I_t; \sf_t} \right)^2}
\\&=
    \sum_{t=1}^{\nt} \E{\est{I_t; \sf_t}^2}
    + \sum_{k \neq l}^{\nt} \E{\est{I_k; \sf_k}\est{I_l; \sf_l}}
\\&=
    \sum_{t=1}^{\nt} \E{\est{I_t; \sf_t}^2}
    + \sum_{k \neq l}^{\nt} \E{\est{I_k}}\E{\est{I_l}},
\end{align*}
where for the last equality we assume independence of the techniques, which usually is a reasonable assumption, and the fact that $\E{\est{I_t; \sf_t}} = \E{\est{I_t}}$, as can easily be seen in the previous unbiasedness computation.

Assuming independence of the acquired samples within a technique, and $\sf_t \in \mathbb{N}_{>0}$, take note of the identity
\begin{align*}
    \E{\left( \sum_{s=1}^{\sf_t} \est{I_t} \right)^2}
    =
    \sf_t \E{\est{I_t}^2} + \sf_t (\sf_t-1) \Esq{\est{I_t}}.
\end{align*}
Using it, we can further simplify the second moment by getting rid of its inherent dependency on $\sf_t$.
For any fixed $t$, the secondary estimator's second moment is of the form
\begin{align*}
    \E{\est{I_t; \sf_t}^2}
&=
    \frac{1}{\sf_t^2}\E{\left( \sum_{s=1}^{r(\sf_t)} \est{I_t} \right)^2}
\\&=
    \frac{\sf_t - \lfloor\sf_t\rfloor}{\sf_t^2}\E{\left( \sum_{s=1}^{\lfloor\sf_t\rfloor+1} \est{I_t} \right)^2}
    \\&\quad+ \frac{1 - \sf_t + \lfloor\sf_t\rfloor}{\sf_t^2}\E{\left( \sum_{s=1}^{\lfloor\sf_t\rfloor} \est{I_t} \right)^2}
\\&=
    \frac{1}{\sf_t} \E{\est{I_t}^2}
    +
    \frac{2\sf_t\lfloor\sf_t\rfloor - \lfloor\sf_t\rfloor^2 - \lfloor\sf_t\rfloor}{\sf_{t}^2}
    \Esq{\est{I_t}}.
\end{align*}
One last trick is to express the first moment squared, $I^2$, in terms as a sum of the technique estimators as follows
\begin{align*}
    I^2
=
    \left(
        \sum_{t=1}^{\nt}
        \E{ \est{I_t} }
    \right)^2
=
    \sum_{t=1}^{\nt}
    \Esq{ \est{I_t} }
    +
    \sum_{k \neq l}^{\nt}
    \E{ \est{I_k} }
    \E{ \est{I_l} }.
\end{align*}
Putting these formulations for the first and second moment together, the formula for the variance of our estimator simply becomes the sum of the variance of all techniques in use
\begin{align*}
    \V{I}
=
    \sum_{t=1}^{\nt}\V{\est{I_t; \sf_t}}
=
    \sum_{t=1}^{\nt}
    \left(
        \frac{1}{\sf_t} \V{\est{I_t}}
    +
        \rho(\sf_t)
        \Esq{\est{I_t}}
    \right),
\end{align*}
where
$\rho(\sf) = \frac{(\sf - \fsf) (\csf - \sf)}{\sf^2}$
arises due to stochastic rounding.

\paragraph{Cost}
Modeling the cost of our estimator is done in a similar way as in previous work \cite{rathEARSEfficiencyawareRussian2022}. We assume the cost $\mathbb{C}_t$ of taking a single sample for technique $t$ to be constant. To obtain the cost of the overall estimator, we multiply each technique's cost by the \emph{real-valued} $\sf_t$ and sum them up. We also include a constant cost term $\mathbb{C}_{\Delta}$ to incorporate all overhead costs. Formally, this yields $\C{\est{I}} = \sum_{t=1}^{\nt} \sf_{t} \mathbb{C}_t + \mathbb{C}_{\Delta}$.

\subsection{Optimizing for Efficiency}

Our goal is to obtain a sample allocation that maximizes efficiency. Instead of doing this directly, we consider the equivalent problem of minimizing \emph{inverse efficiency} $\mathcal{E}^{-1}$ to improve the readability and reduce clutter of the following equations. At a minimum, the derivative necessarily has a root. Therefore, we first compute the derivative w.r.t. all $\sf_t$ of inverse efficiency in this section. In the next sections, we then find an expression for its minimum.

\paragraph{Efficiency derivative.}
We can use the product formula to get
\begin{align*}
    \diff[{\mathcal{E}^{-1}[\est{I}]}]{\sf_{t}}
    =
    \diff[\V{\est{I}}]{\sf_{t}}\,\C{\est{I}} + \V{\est{I}} \diff[\C{\est{I}}]{\sf_{t}},
\end{align*}
revealing that we need derivatives of the cost and variance expressions computed previously.

\paragraph{Cost derivative}
The derivative of cost is easy as only one of the summands depends on $\sf_t$.
Therefore: $\diff[\C{\est{I}}]{\sf_t} = \mathbb{C}_{t}$.

\paragraph{Variance derivative}
The derivative of variance is given by
\begin{align*}
    \diff[\V{\est{I}}]{\sf_{t}}
&=
    \sum_{k=1}^{\nt}
    \left(
        \diff[ ]{\sf_{t}} \frac{1}{\sf_k} \V{\est{I_k}}
    +
        \diff[ ]{\sf_{t}}
        \rho(\sf_k)
        \Esq{\est{I_k}}
    \right)
\\&=
    - \frac{1}{\sf_t^2} \V{\est{I_t}}
    +
        \diff[\rho(\sf_t)]{\sf_{t}}
        \Esq{\est{I_t}}
    \\&\quad+
    \sum_{k=1}^{\nt}
    \underbrace{
        \left(
            \frac{1}{\sf_k} \diff[\V{\est{I_k}}]{\sf_{t}}
        +
            \rho(\sf_k)
            \diff[\Esq{\est{I_k}}]{\sf_{t}}
        \right)
    }_{=0,\text{ for budget-unaware MIS weights}}
\end{align*}
Note that, at this point, the sum over all $n_t$ techniques is 0 if the MIS weights are budget-unaware, i.e., do not contain $\sf$.
Moreover, the derivative of $\rho$ in this equation, which arises due to stochastic rounding, forces us to make a simplification that has been done by previous work in similar situations as well \cite{rathEARSEfficiencyawareRussian2022}, \ie
we assume $\rho\left(\sf\right) = \max(1, \frac{1}{\sf}) - 1$.
It is easy to verify that this simplification only affects the $\sf > 1$ case, giving us
\begin{equation}
    \V{\est{I_t; \beta_t}}
    = \begin{cases}
        \frac{1}{\sf_t} \E{\est{I_t}^2} -
            \Esq{\est{I_t}}
        & \text{if}\;\sf_t < 1\\
        \frac{1}{\sf_t} \V{\est{I_t}} & \text{otherwise}
    \end{cases}
\end{equation}
as a formulation for variance, and
\begin{align*}
    \diff[\V{\est{I}}]{\sf_{t}}
&=
    \begin{cases}
        - \frac{1}{\sf_t^2} \E{\est{I_t}^2}
        & \text{if}\;\sf_t < 1,\\
        - \frac{1}{\sf_t^2} \V{\est{I_t}}
        & \text{otherwise}
    \end{cases}
\end{align*}
for the derivative
in the \emph{budget-unaware} scenario.

In the budget-aware scenario, the exact expression strongly depends on the MIS weight heuristic, but is not generally convex anymore even for simple heuristics like the balance heuristic.
Hence, we will focus on the budget-unaware case and treat it as a simplification for the budget-aware case. With this simplification, we lose optimality, but we show empirically that it performs well both in simple 1D examples (see \cref{fig:1d_eval}) and complex rendering applications (see paper).

Putting the equations together, we obtain
\begin{align}\label{eq:ieff-deriv}
    \diff[{\mathcal{E}^{-1}[\est{I}]}]{\sf_{t}}
=
    \begin{cases}
        - \frac{1}{\sf_t^2} \E{\est{I_t}^2} \C{\est{I}} + \V{\est{I}} \mathbb{C}_t
        & \text{if}\;\sf_t \leq 1,\\
        - \frac{1}{\sf_t^2} \V{\est{I_t}} \C{\est{I}} + \V{\est{I}} \mathbb{C}_t
        & \text{otherw}.
    \end{cases}
\end{align}

\paragraph{Budget-aware derivatives}
The variance's derivative includes some extra terms that are generated by the derivative of the MIS weights. For any fixed $k$, these extra terms take the form
\begin{align*}
    &\quad\frac{1}{\sf_k} \diff[\V{\est{I_k}}]{\sf_{t}}
    +
    \rho(\sf_k)
    \diff[\Esq{\est{I_k}}]{\sf_{t}}
\\&=
    \frac{2}{\sf_k} \E{\est{I_k}\diff[\est{I_k}]{\sf_{t}}}
    +
    \frac{2}{\sf_k} \E{\est{I_k}} \Esq{\diff[\est{I_k}]{\sf_{t}}}
    \\&\quad+
    2
    \rho(\sf_k)
    \E{\est{I_k}} \Esq{\diff[\est{I_k}]{\sf_{t}}}
\\&=
    \frac{2}{\sf_k} \E{\est{I_k}
        \sum_{i=1}^{\ni} \frac{f_i(x)}{p_k(x)} \diff[w_{ik}(x)]{\sf_{t}}
    }
    \\&\quad
    +
    \frac{2}{\sf_k} \E{\est{I_k}} \Esq{
        \sum_{i=1}^{\ni} \frac{f_i(x)}{p_k(x)} \diff[w_{ik}(x)]{\sf_{t}}
    }
\\&\quad+
    2
    \rho(\sf_k)
    \E{\est{I_k}} \Esq{
        \sum_{i=1}^{\ni} \frac{f_i(x)}{p_k(x)} \diff[w_{ik}(x)]{\sf_{t}}
    } .
\end{align*}

\subsection{Fixed-Point Iteration}
To devise a fixed-point scheme, we need to find the inverse efficiency's derivative's roots, which we achieve by rearranging \cref{eq:ieff-deriv}.
For brevity, let $M_t = \diff[\V{\est{I}}]{\sf_{t}}$. Then
\begin{align*}
    &\diff[{\mathcal{E}^{-1}[\est{I}]}]{\sf_{t}} \overset{!}{=} 0
\\\iff&
    - \frac{1}{\sf_t^2} M_t \C{\est{I}} + \V{\est{I}} \mathbb{C}_t = 0
\\\iff&
    \sf_t = \sqrt{\frac{M_t}{\V{\est{I}}} \frac{\C{\est{I}}}{\mathbb{C}_t}},
\end{align*}
where the last equivalence holds because $\sf_t > 0$.

With this equation, there is to note that $\sf_t$ appears on both sides. And non-linearly so on the right-hand side as a part of $\V{\est{I}}$. Hence, an analytical solution becomes impractical and we have to resort to numerical methods. The equation looks similar to what previous work \cite{rathEARSEfficiencyawareRussian2022} has encountered before and solved with a fixed-point iteration scheme.
Following their example, we can employ a fixed-point iteration by using the equation above as an update function and compute the new sample allocations based on the variances and costs of using the previous iterations' allocations, as given by
\begin{align*}
    \sf_{t,i+1}
=
    \begin{cases}
        \sf_{t,i}^{\text{RR}} = \sqrt{\frac{\E{\est{I_t; \sf_{t,i}}^2}}{\V{\est{I; \bm{\sf}_{i}}}} \frac{\C{\est{I; \bm{\sf}_{i}}}}{\mathbb{C}_t}}
        & \text{if}\;\sf_{t,i}^{\text{RR}} < 1,\\
        \sf_{t,i}^\text{S\phantom{...}} = \sqrt{\frac{\V{\est{I_t; \sf_{t,i}}}}{\V{\est{I; \bm{\sf}_{i}}}} \frac{\C{\est{I; \bm{\sf}_{i}}}}{\mathbb{C}_t}}
        & \text{if}\;\sf_{t,i}^\text{S\phantom{...}} > 1,\\
        1
        & \text{otherwise},\\
    \end{cases}
\end{align*}
where $\bm{\sf}_{i}$ is the vector of all $\sf_{t,i}$ of the $i$-th iteration.
The third case arises due to a technicality. With our proxy, the problem is split into two cases, one where $\sf_t \leq 1$ and one where $\sf_t > 1$.
In some scenarios, an update step using the first case would lead to $\sf_{t, i+1} \geq 1$, whereas an update step in the second case would lead to $\sf_{t, i+1} \leq 1$. The two cases are in disagreement. Therefore, in such scenarios, we resort to following neither decision by allocating a single sample.

\subsection{Proof of Optimality for Budget-Unaware Weights}
We know that the efficiency is of the form
\begin{equation*}
    \mathcal{E}^{-1}(\mathbb{\sf}) = \underbrace{\left(
        \sum_t^{\nt} \frac{1}{\sf_t} \mathbb{V}[\est{I_t}] + \mathbb{V}_\Delta
    \right)}_{\mathbb{V}_\Sigma} \underbrace{\left(
        \sum_t^{\nt} \sf_t \mathbb{C}[\est{I_t}] + \mathbb{C}_\Delta
    \right)}_{\mathbb{C}_\Sigma} ,
\end{equation*}
and hence optimal $\sf_t^*$ must satisfy
\begin{equation*}
    \sf_t^* = \sqrt{\frac{
        \mathbb{V}[\est{I_t}]
    }{
        \mathbb{C}[\est{I_t}]
    } \frac{
        \mathbb{C}_\Sigma
    }{
        \mathbb{V}_\Sigma
    }} = \lambda\,\sqrt{\frac{
        \mathbb{V}[\est{I_t}]
    }{
        \mathbb{C}[\est{I_t}]
    }} .
\end{equation*}
For budget-unaware weights, the variance $\mathbb{V}[\est{I_t}]$ is independent of $\sf_t$, and hence all that remains to be found is $\lambda$. Rewriting the inverse efficiency in terms of $\lambda$ yields
\begin{equation*}
    \mathcal{E}^{-1}(\lambda) = \left(
        \frac{1}{\lambda} M
        + \mathbb{V}_\Delta
    \right) \left(
        \lambda M
        + \mathbb{C}_\Delta
    \right),
\end{equation*}
where $M = \nt \sqrt{\mathbb{V}[\est{I_t}] \mathbb{C}[\est{I_t}]}$. Since our update formula enforces that we move on the line defined by $\lambda$, our fixed-point scheme degenerates to the EARS fixed-point scheme after a single step, which has been proven to be optimal \cite{rathEARSEfficiencyawareRussian2022}.

\section{Additional discussion}

\subsection{Rounding}
\begin{figure*}[h]
    \centering
    \includegraphics[width=\textwidth]{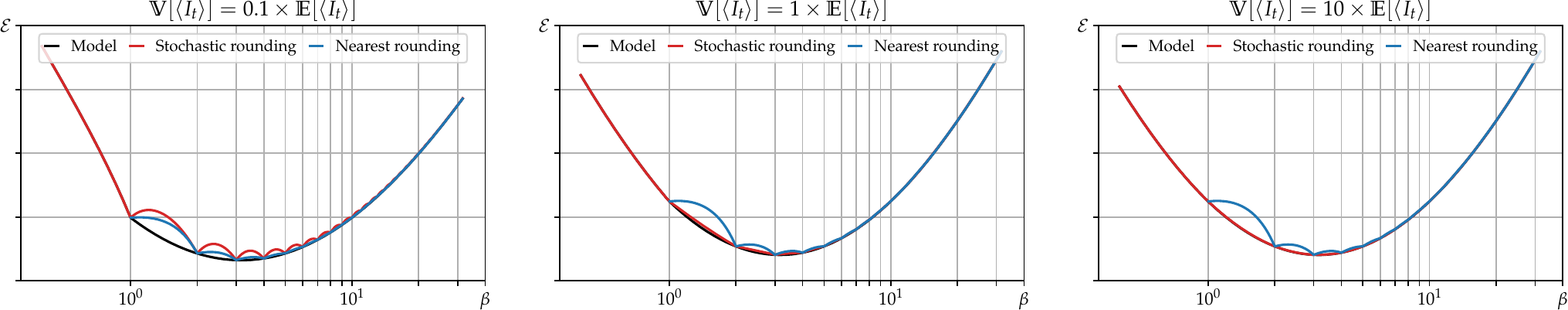}
    \captionsetup[subfigure]{aboveskip=-7pt}
    \begin{minipage}[t]{.33\textwidth}
        \subcaption{Low variance}
    \end{minipage}%
    \begin{minipage}[t]{.33\textwidth}
        \subcaption{Medium variance}
    \end{minipage}%
    \begin{minipage}[t]{.33\textwidth}
        \subcaption{High variance}
    \end{minipage}%
    \caption{We compare the efficiency of stochastic rounding, rounding to nearest integers, and the simplified model for the secondary estimator variance our work has in common with previous works. The performance of stochastic rounding depends on the ratio of estimator variance to expected value. All models perform the same for RR ($\beta < 1$) and at integer splitting factors. \textbf{(a)} For low variance estimators, stochastic rounding can introduce more noise than nearest rounding. \textbf{(b)} When the estimator variance is on the same order of magnitude as its expected value, stochastic rounding becomes the best option. \textbf{(c)} Stochastic rounding and the simplified model begin to agree perfectly when the variance is high.}
    \Description{}
    \label{fig:rounding}
\end{figure*}

\paragraph{Rounding variants}
Rounding fractional sample counts to integers, which is required since we can only take an integer amount of samples, always introduces additional noise in the secondary estimator. The shape and magnitude of noise depend on how the secondary estimator is normalized after rounding. We evaluate two unbiased options: \textbf{(1)}~dividing by the real-valued sample count as done by previous works \cite{vorbaAdjointdrivenRussianRoulette2016, rathEARSEfficiencyawareRussian2022}, and \textbf{(2)}~dividing by the rounded sample count in the case of splitting and real-valued count for RR. The theoretical difference in noise is demonstrated in \cref{fig:rounding}.
In practice, we have found that both techniques perform equally well, and hence for consistency with previous works, we normalize by real-valued sample counts in all of our evaluations.

\paragraph{Low discrepancy rounding}
In practice, we apply a small trick to reduce the variance introduced by stochastic rounding by roughly 2\%. With naïve stochastic rounding, every technique draws an independent random number to decide whether to round the number of samples up or down. Looking at the sampling process as a whole, the decision is not whether 1 sample is taken or not, but rather whether $\nt$ samples are taken, each individually. With an unlucky draw none of the allocations are rounded up, whereas with a lucky draw, all of them are rounded up. We instead only draw a \emph{single} random number to decide the rounding of \emph{all} techniques (see alg. \ref{alg:low-disc-stoch-round}).
After a technique was rounded, the probability of rounding up the next sample increases or decreases with whether the previous ones were rounded down or up, respectively. Overall, this leads to the random process only deciding between taking 1 sample more or less, instead of $\nt$ samples, which reduces variance slightly.

\begin{algorithm}
\caption{Low Discrepancy Stochastic Rounding}\label{alg:low-disc-stoch-round}
    \tcc{Given stochastic number of samples $\sf_t\in\mathbb{R}^+_0$}
    $r \gets random()$ \Comment*[r]{Uniform random number in $[0,1]$}
    \For(\tcc*[f]{Round all $\nt$ sample allocations}){$t = 1$ \KwTo $\nt$}{
        $\gamma_t \gets \lfloor \sf_t + r \rfloor$ \Comment*[r]{$\sf_t\in\mathbb{R}^+_0$; stochast. rounded $\gamma_t\in\mathbb{N}$}
        $r \gets r + \sf_t - \gamma_t$\ \Comment*[r]{Remaining fractional random part}
    }
    \tcc{Use each $\gamma_t$ as number of samples for technique $t$.}
\end{algorithm}

\subsection{Convergence}
In addition to the rendered images that show an improvement over state-of-the-art methods and the visualizations of where which technique is used, we provide efficiency and sample allocation plots over rendering time. In \cref{fig:convergence_plot}, we present efficiency and sample allocation plots of a small selection of scenes in path guiding as well as in bidirectional path tracing. We can observe the curves flattening, or being near constant at the end of the rendering process, indicating a sufficiently converged sample-allocation fixed point in our proxy model.

\begin{figure*}
    \centering
    \vspace{5mm}
    Path guiding\vspace{2mm}\\
    \includegraphics[width=\textwidth,interpolate=false]{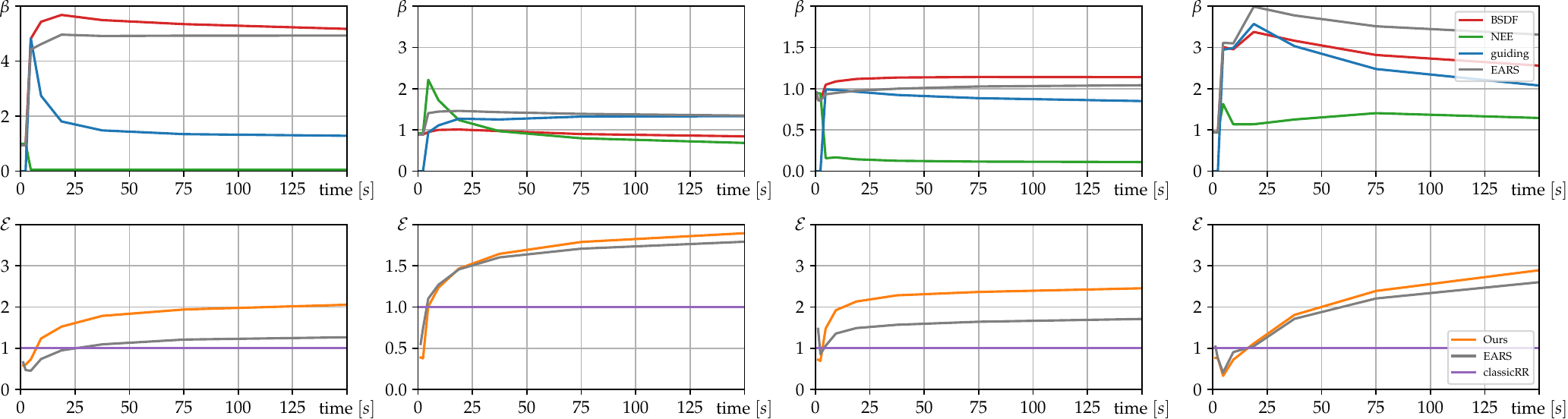}
    \vspace{1mm}\\
    Bidirectional path tracing\vspace{2mm}\\
    \includegraphics[width=\textwidth,interpolate=false]{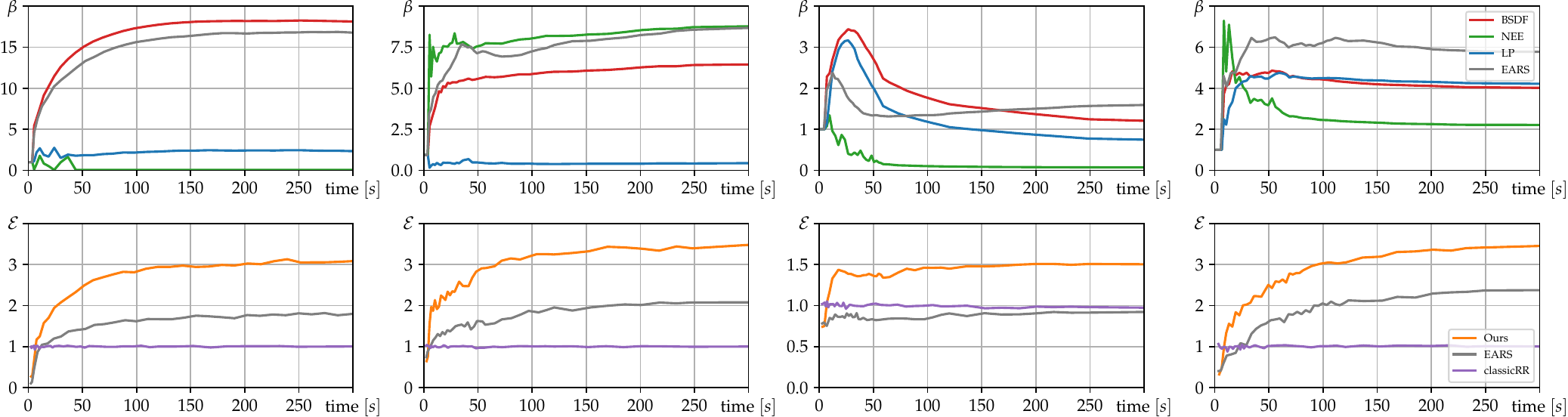}
    \captionsetup[subfigure]{aboveskip=1pt}
    \begin{minipage}[t]{.245\textwidth}
        \subcaption{\textsc{Glossy Kitchen}}
    \end{minipage}%
    \begin{minipage}[t]{.245\textwidth}
        \subcaption{\textsc{Dining Room}}
    \end{minipage}%
    \begin{minipage}[t]{.245\textwidth}
        \subcaption{\textsc{Glossy Bathroom}}
    \end{minipage}%
    \begin{minipage}[t]{.245\textwidth}
        \subcaption{\textsc{Bookshelf}}
    \end{minipage}%
    \caption{
    We show convergence of the average primary hit point sample allocation (first row) and overall efficiency (second row, higher is better) in four selected scenes for path guiding and BDPT. The evaluation setup is the same as in the main text. Our fixed-point scheme quickly converges towards solutions that are better than the baseline and previous state-of-the-art.
    }
    \label{fig:convergence_plot}
    \Description{}
\end{figure*}

\section{Additional Results}
\subsection{1D Examples}
Apart from verifying the adequacy of our proxy model in complex rendering applications, we also investigate its performance and deviation from the true efficiency in a number of 1D examples visualized in \cref{fig:1d_eval}.
While our proxy model matches exactly for the budget-unaware case, its optimum can deviate from the true optimal solution for budget-aware weights.
This is most pronounced when the optimal solution is to discard all but one distribution, either because one strategy samples perfectly (second last row) or the problem becomes non-convex as individual strategies perform better than the MIS combination (last row).

\begin{figure*}[p]
  \centering
  \includegraphics[width=\linewidth]{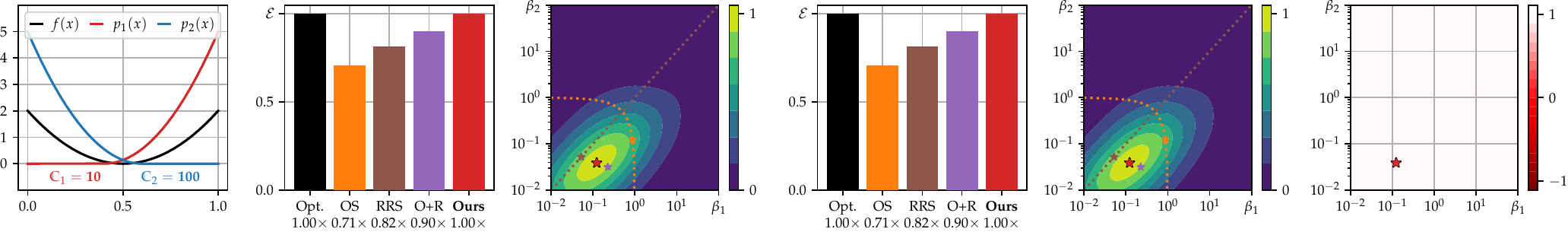} 
  \includegraphics[width=\linewidth]{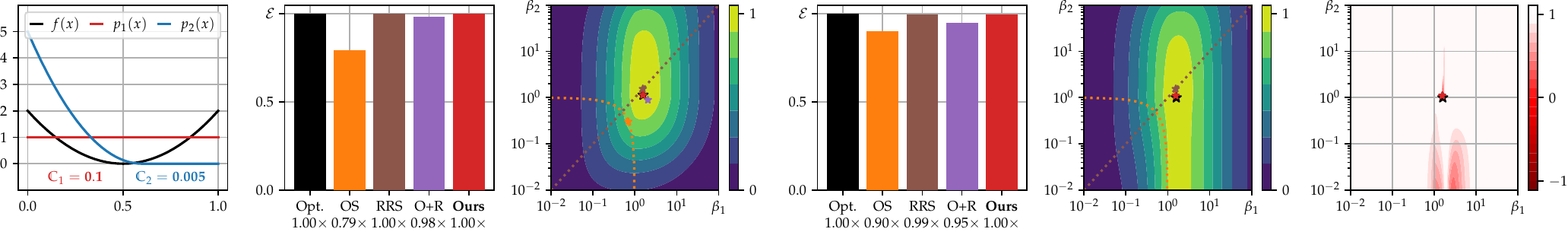} 
  \includegraphics[width=\linewidth]{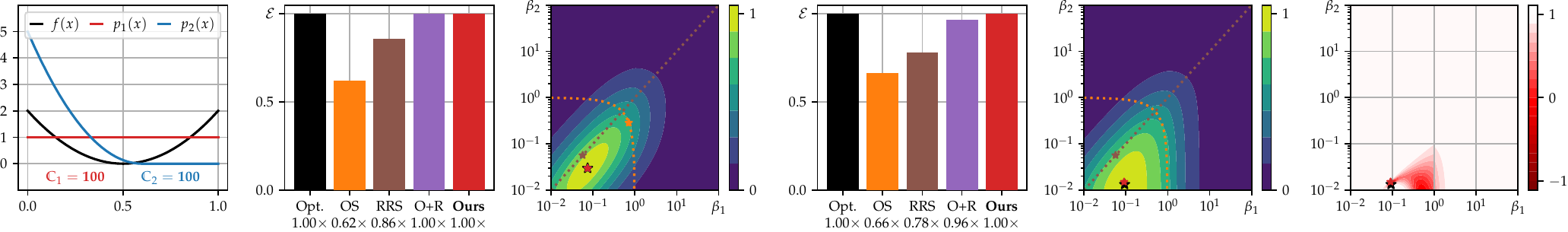} 
  \includegraphics[width=\linewidth]{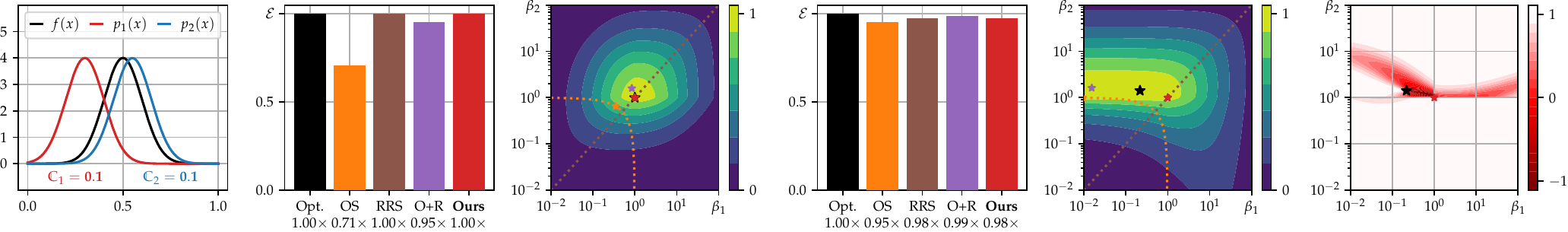} 
  \includegraphics[width=\linewidth]{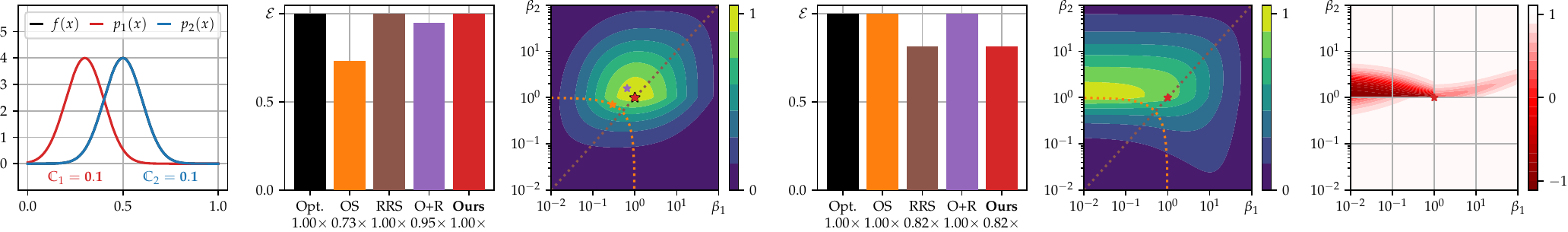} 
  \includegraphics[width=\linewidth]{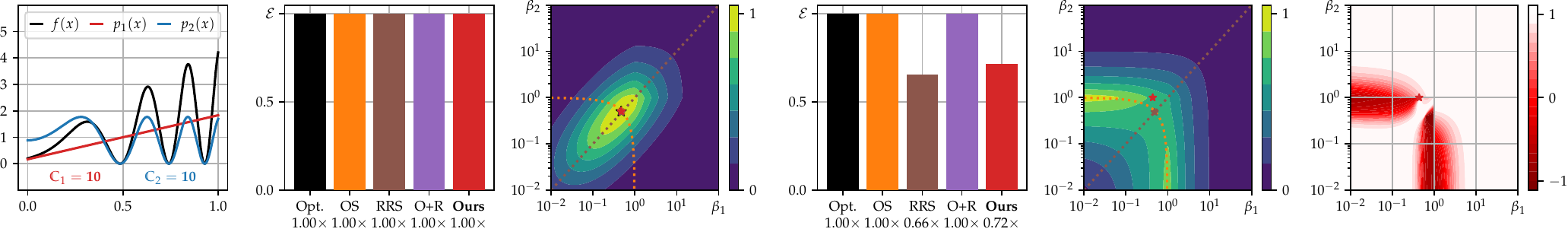} 
  \captionsetup[subfigure]{aboveskip=-7pt}
  \begin{minipage}[t]{.16\textwidth}
      \subcaption{Integrand}
  \end{minipage}%
  \begin{minipage}[t]{.33\textwidth}
      \subcaption{Budget-unaware balance heuristic}
  \end{minipage}
  \begin{minipage}[t]{.33\textwidth}
      \subcaption{Budget-aware balance heuristic}
  \end{minipage}
  \begin{minipage}[t]{.16\textwidth}
      \subcaption{Proxy error}
  \end{minipage}%
  \caption{
      \textbf{(a)} We evaluate the performance of our model on simple 1D examples, in which a single integrand $f$ is estimated by two techniques $p_1$ and $p_2$, the costs of which are indicated by $\mathbb{C}_1$ and $\mathbb{C}_2$ ($\mathbb{C}_\Delta = 1$ and $\mathbb{V}_\Delta = 1$ across all examples).
      \textbf{(b)} We compare the efficiency of four approaches: Optimal mixture sampling (``OS''), optimal Russian roulette and splitting (``RRS''), combining the two (``O+R''), and our proxy model (``Ours''). Note that combining optimal mixture sampling and RRS does not yield optimal performance. Our model optimizes ratios and total sample counts jointly, resulting in optimal performance for budget-unaware MIS weights.
      \textbf{(c)} For budget-aware MIS weights, the optimum predicted by our model can deviate from the true efficiency optimum.
      \textbf{(d)} We investigate the error of our proxy by plotting the dot product of the true gradients and our proxy gradients (``1'' indicates a perfect match, and ``-1'' indicates opposing directions).
  }
  \label{fig:1d_eval}
  \Description{}
\end{figure*}

\subsection{Mixture Optimization in Path Guiding}
In path guiding, we optimize the BSDF/guiding ratio of previous work using the KL divergence extended to image contributions \cite{rath2020variance}. This cannot directly maximize efficiency because \textbf{(1)} it does not contain information about cost, \textbf{(2)} only optimizes decisions locally rather than balancing variance across pixels and regions, and \textbf{(3)} optimizes KL divergence instead of variance. While \textbf{(1)} and \textbf{(2)} are fundamental limitations of the approach, \textbf{(3)} is in theory addressed by the variance optimizing variant \cite{vorbaPathGuidingProduction2019}, but performed poorly across our tests scenes--likely due to the Adam optimizer drastically reducing its step-size in the presence of strong outliers. Hence, as suggested by the original authors, we use the KL divergence instead.
A brief selection of scenes rendered for 5 minutes showcasing minimizing for \textbf{(a)}~KL~divergence, \textbf{(b)}~KL~divergence with image contributions, \textbf{(c)} variance, and \textbf{(d)} variance with image contributions can be found in \cref{fig:KL}.
\begin{minipage}{\textwidth}
    \includegraphics[width=\textwidth,interpolate=false]{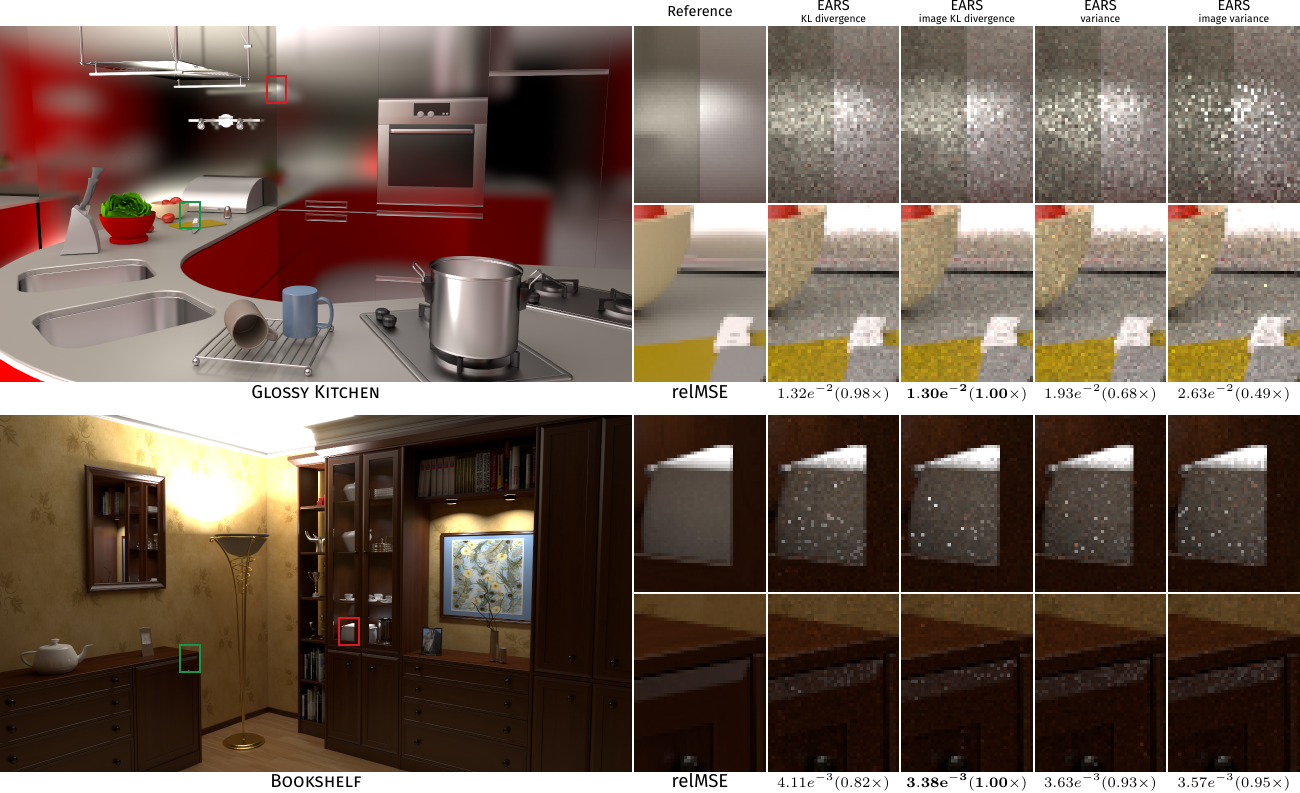}
    \captionof{figure}{
    We compare variants of gradient descent for the BSDF and guiding sample mixture on top of EARS in guiding:
    \textbf{(a)} KL divergence,
    \textbf{(b)} KL divergence of image contributions,
    \textbf{(c)} variance, and
    \textbf{(d)} variance of image contributions.
    Across our set of test scenes, KL divergence with image contributions performs best.}
    \label{fig:KL}
    \Description{}
\end{minipage}

\subsection{Additional Renders}
For brevity, we have only included the most interesting scenes in our main text. We show results for all other scenes from our test set in \cref{fig:GEARS-evaluation-supplemental} (for path guiding) and \cref{fig:BEARS-evaluation-supplemental} (for BDPT). As with the scenes shown in the paper, our method consistently outperforms the respective previous state-of-the-art.

\begin{figure*}
    \includegraphics[width=.951\textwidth,interpolate=false]{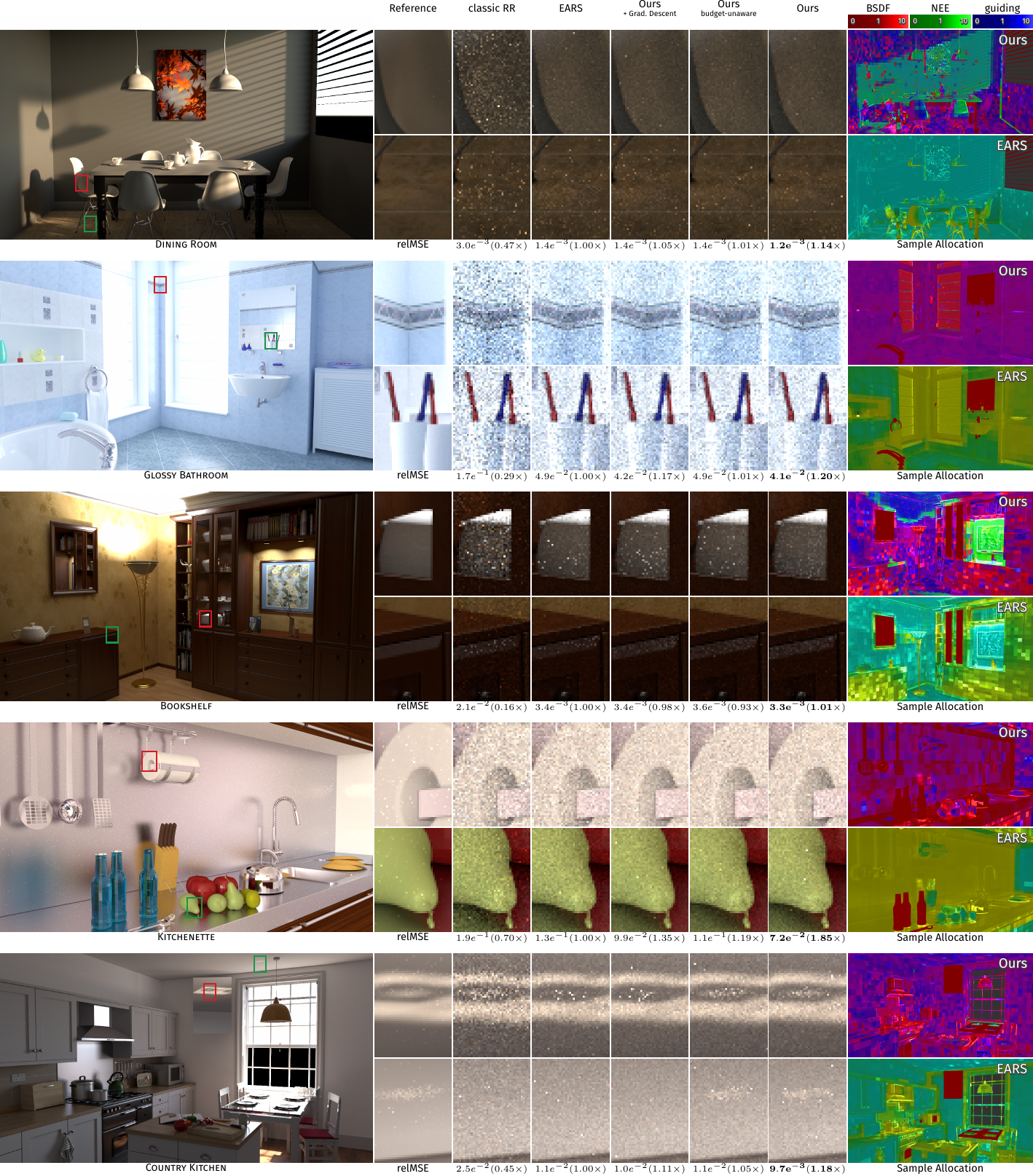}
    \caption{
        We compare our method against state-of-the-art EARS applied to guiding in the 5 scenes not provided in the paper. The numbers below the crops correspond to the relative mean squared error (relMSE, lower is better), with the speedup compared against EARS in parentheses (higher is better). To the right, we visualize the allocation decisions EARS and our method take. The color channels red, green and blue correspond to BSDF, NEE, and guiding samples, respectively, with brighter colors indicating higher sample counts. The other 5 scenes we tested are provided in the main text.
        Our ``+ Grad. Descent''
        test uses gradient descent for the BSDF/guiding mixture and our method to combine it with NEE using budget-aware weights.
    }
    \label{fig:GEARS-evaluation-supplemental}
    \Description{}
\end{figure*}

\begin{figure*}
    \includegraphics[width=.951\textwidth,interpolate=false]{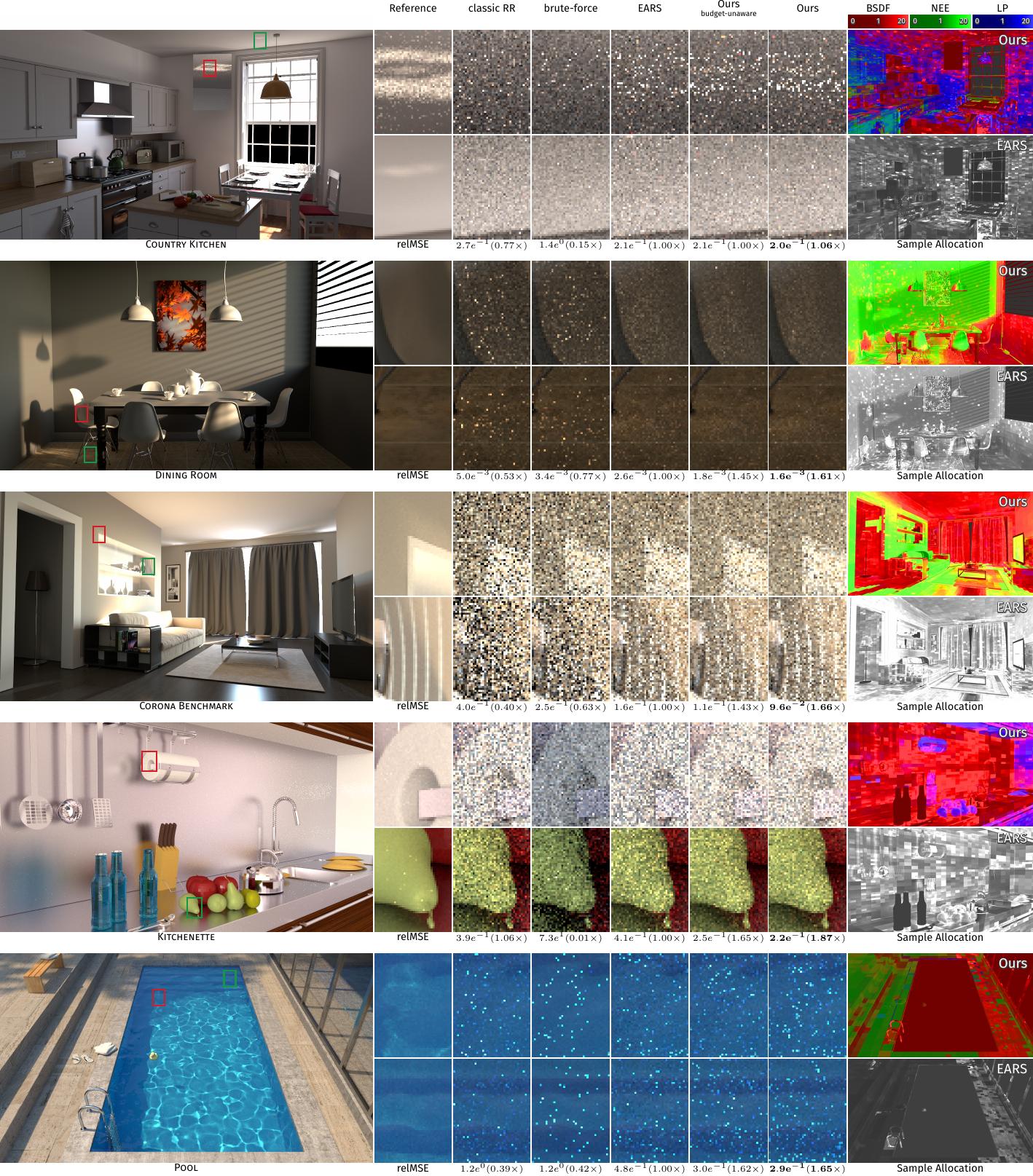}
    \caption{
        We compare our method against state-of-the-art EARS applied to BDPT in the 5 scenes not provided in the paper in a 5 minute equal-time comparison. The numbers below the crops correspond to the relative mean squared error (relMSE, lower is better), with the speedup compared against EARS in parentheses (higher is better). To the right, we visualize the allocation decisions EARS and our method take. The color channels red, green and blue correspond to BSDF, NEE, and light path samples, respectively, with brighter colors indicating higher sample counts. The other 5 scenes we tested are provided in the main text.
    }
    \label{fig:BEARS-evaluation-supplemental}
    \Description{}
\end{figure*}

\fi
\end{document}